\documentclass[onecolumn,showpacs]{revtex4}

\usepackage{graphicx}% Include figure files
\usepackage{dcolumn}% Align table columns on decimal point
\usepackage{bm}% bold math

\input epsf

\begin{document}

\title{\Large Dynamics of Logamediate and Intermediate Scenarios in the Dark Energy Filled Universe}

\author{\bf Piyali Bagchi Khatua$^1$\footnote{piyali.bagchi@yahoo.co.in}
and Ujjal Debnath$^2$\footnote{ujjaldebnath@yahoo.com} }

\affiliation{$^1$Department of Computer Science and Engineering,
Netaji Subhas Engineering College, Garia, Kolkata-700 152, India.\\
$^2$Department of Mathematics, Bengal Engineering and Science
University, Shibpur, Howrah-711 103, India. }

\date{\today}

\begin{abstract}
We have considered a model of two component mixture i.e., mixture
of Chaplygin gas and barotropic fluid with tachyonic field. In the
case, when they have no interaction then both of them retain their
own properties. Let us consider an energy flow between barotropic
and tachyonic fluids. In both the cases we find the exact
solutions for the tachyonic field and the tachyonic potential and
show that the tachyonic potential follows the asymptotic behavior.
We have considered an interaction between these two fluids by
introducing a coupling term. Finally, we have considered a model
of three component mixture i.e., mixture of tachyonic field,
Chaplygin gas and barotropic fluid with or without interaction.
The coupling functions decays with time indicating a strong energy
flow at the initial period and weak stable interaction at later
stage. To keep the observational support of recent acceleration we
have considered two particular forms (i) Logamediate Scenario and
(ii) Intermediate Scenario, of evolution of the Universe. We have
examined the natures of the recent developed statefinder
parameters and slow-roll parameters in both scenarios with and
without interactions in whole evolution of the universe.\\
\end{abstract}


\maketitle

\section{\normalsize\bf{Introduction}}

Recent observations of the luminosity of type Ia supernovae
indicate [1-7] an accelerated expansion of the universe and lead
to the search for a new type of matter which violate the strong
energy condition $\rho+3p<0$. The matter consent responsible for
such a condition to be satisfied at a certain stage of evolution
of the universe is referred to as {\it dark energy}. There are
different candidates to play the role of the dark energy. The type
of dark energy represented by a scalar field is often called {\it
Quintessence}. The transition from a universe filled with matter
to an exponentially expanding universe does not necessarily
require the presence of the scalar field as the only alternative.
In particular one can try another alternative by using an exotic
type of fluid - the so-called Chaplygin gas [8-14]. Assume that
the cosmological model, which is denoted by $\Lambda$CDM, contains
a cosmological constant $\Lambda$ and the cold dark matter. In the
presence of an interaction the dark energy can achieve a stable
equilibrium that differs from the usual de Sitter case. The
effective equations of state of matter and dark energy coincide
and behave like cold dark matter (CDM) at early times. Actually,
dark energy is a mysterious fluid, contains enough negative
pressure causes the present day acceleration.\\

The energy-momentum tensor of the tachyonic field [15] can be seen
as a combination of two fluids, dust with pressure zero and a
cosmological constant with $p=-\rho$, thus generating enough
negative pressure such as to drive acceleration. Also the
tachyonic field has a potential which has an unstable maximum at
the origin and decays to almost zero as the field goes to
infinity. Depending on various forms of this potential following
this asymptotic behaviour a lot of works have been carried out on
tachyonic dark energy [16-19], tachyonic dark matter [20-22] and
inflation models [23,24].\\

Here we consider a model of two component mixture i.e., mixture of
Chaplygin gas and barotropic fluid with tachyonic field. In the
case, when they have no interaction then both of them retain their
own properties. Let us consider an energy flow between barotropic
and tachyonic fluids. In both the cases we find the exact
solutions for the tachyonic field and the tachyonic potential and
show that the tachyonic potential follows the asymptotic behavior.
Later we have also considered an interaction between these two
fluids by introducing a coupling term. The coupling function
decays with time indicating a strong energy flow at the initial
period and weak stable interaction at later stage. To keep the
observational support of recent acceleration we have considered
two particular forms: (i) Logamediate Scenario [25] and (ii)
Intermediate Scenario[25, 26], of evolution of the Universe. The
intermediate and logamediate Scenarios are motivated by
considering a class of possible cosmological solutions with
indefinite expansion which result from imposing weak general
conditions on the cosmological model. The intermediate Scenario
satisfies the bounds on the spectral index $n_{s}$ and ratio of
tensor-to-scalar perturbations, $r$, as measured by the latest
observations of the CMB. For observationally viable models of
logamediate Scenario, the ratio of tensor-to-scalar perturbations,
$r$, must be small and that the power spectrum can be either red
or blue tilted, depending on the specific parameters of the model.
It has the interesting property that the cooperative evolution We
have examined the nature of the recent developed statefinder
parameters [27] and slow-roll
parameters [25] in whole evolution of the universe.\\

The paper is organized as follows: Section II deals with the field
equations of the tachyonic field in logamediate and intermediate
scenarios of the universe. In sections III we have considered
models represented by mixture of tachyonic field with GCG. In
sections IV we have considered models represented by mixture of
tachyonic field with barotropic fluid. In sections V we have
considered models represented by mixture of tachyonic field with
GCG and Barotropic fluid. These three sections are each subdivided
into two parts showing the effect of these models with or without
interaction. We have found also the expressions of
slow-roll-parameter. We have taken some particular values of the
parameters and constants for the graphical representation.
The paper ends with a short discussion in section VI.\\

\section{\normalsize\bf{Einstein Field Equations and Tachyonic Fluid
Model}}

The metric of a spatially flat isotropic and homogeneous Universe
in FRW model is
\begin{equation}
ds^{2}=dt^{2}-a^{2}(t)\left[dr^{2}+r^{2}(d\theta^{2}+sin^{2}\theta
d\phi^{2})\right]
\end{equation}
where $a(t)$ is the scale factor of the universe. The Einstein
field equations are (choosing $8\pi G = c = 1$)
\begin{equation}
3H^2 = \rho_{tot}
\end{equation}
and
\begin{equation}
6(\dot{H}+H^2) = -(\rho_{tot}+p_{tot})
\end{equation}
where, $\rho_{tot}$ and $p_{tot}$ are respectively the total
energy density and the pressure of the Universe. Here $H$ is
called Hubble parameter defined as,
\begin{equation}
 H = \frac{\dot{a}}{a}
\end{equation}

In the following, we'll discuss the natures of statefinder
parameters and deceleration parameter in the particular forms of
logamediate and intermediate
Scenario.\\

\subsection{\normalsize\bf{Logamediate Scenario}}
Consider a particular form of Logamediate Scenario [25], where the
form of the scale factor $a(t)$ is defined as,
\begin{equation}
a(t)=\exp (A(\ln t)^{\alpha})
\end{equation}
where $A \alpha>0$ and $\alpha>1$. When $\alpha=1$, this model
reduces to power-law form. The logamediate form is motivated by
considering a class of possible cosmological solutions with
indefinite expansion which result from imposing weak general
conditions on the cosmological model. Barrow [25] has found in
their model, the observational ranges of the parameters are as
follows: $1.5\times 10^{-92}\le A \le 2.1\times 10^{-2}$ and $2\le
\alpha \le 50$. The Hubble parameter $H=\frac{\dot{a}}{a}$
becomes,
\begin{equation}
H=\frac{A\alpha}{t}(\ln t)^{\alpha-1}
\end{equation}
Hence from (6) we get,
\begin{equation}
\frac{\dot H}{H}=\frac{\alpha-1-\ln t}{t\ln t}
\end{equation}
and
\begin{equation}
\frac{\ddot H}{H}=\frac{2(\ln t)^2-3(\alpha-1)\ln
t+(\alpha-1)(\alpha-2)}{t^2 (\ln t)^2}
\end{equation}

Putting the value of $a(t)$ in the deceleration parameter
$q=-\frac{a \ddot{a}}{\dot{a}^2}$ we get,
\begin{equation}
q=-1+\frac{\ln t-\alpha+1}{A\alpha (\ln{t})^\alpha}
\end{equation}
where $a(t)$ is the scale factor. Fig.1 represents the variation
of $H$ against $q$ for different values of $\alpha$.\\

\begin{figure}
\includegraphics[height=2in]{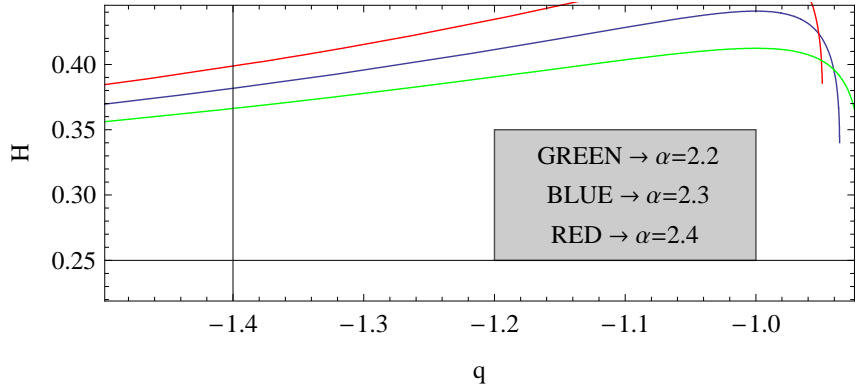}~~~~
\caption{The variation of $H$ against $q$ for logamediate Scenario
with $A = 1$ and $\alpha=2.2,2.3,2.4$} \vspace{7mm}
\end{figure}

The flat Friedmann model which is analyzed in terms of the
statefinder parameters. The trajectories in the $\{s, r\}$ plane
of different cosmological models shows different behavior. The
statefinder diagnostic of SNAP observations used to discriminate
between different dark energy models. The statefinder diagnostic
pair is constructed from the scale factor $a(t)$. The statefinder
diagnostic pair is denoted as $\{s,r\}$ and defined as [27],
\begin{equation}
 r=\frac{\dddot{a}}{a H^3}
 ~~~\text{and}~~~
s=\frac{r-1}{3(q-\frac{1}{2})}
\end{equation}

From (5), (6), (9) and (10) we get,
\begin{equation}
r=1+\frac{3(\alpha-1)}{A\alpha (\ln{t})^\alpha}-\frac{3}{A\alpha
(\ln{t})^{\alpha-1}}+\frac{2}{A^{2}\alpha^{2}(\ln{t})^{2\alpha-2}}
-\frac{3(\alpha-1)}{A^{2}\alpha^{2}(\ln{t})^{2\alpha-1}}
+\frac{(\alpha-1)(\alpha-2)}{A^2\alpha^2 (\ln{t})^{2\alpha}}
\end{equation}
and
\begin{equation}
s=\frac{\frac{3(\alpha-1)}{A\alpha(\ln{t})^{\alpha}}-\frac{3}{A\alpha
(\ln{t})^{\alpha-1}}+\frac{2}{A^{2}\alpha^{2}(\ln{t})^{2\alpha-2}}
-\frac{3(\alpha-1)}{A^{2}\alpha^{2}(\ln{t})^{2\alpha-1}}
+\frac{(\alpha-1)(\alpha-2)}{A^{2}\alpha^{2}(\ln{t})^{2\alpha}}}
{\frac{3}{A\alpha(\ln{t})^{\alpha-1}}-\frac{3(\alpha-1)}{A\alpha(\ln{t})^{\alpha}}-\frac{9}{2}}
\end{equation}

Fig.2 represents the variation of $s$ against $r$ for different
values of $\alpha$. We see that at first $r$ increases with $s$
decreases and then $r$ decreases with increasing $s$. Here we see
that $r$ restricts always positive value upto some stage and may
be takes negative value at final stage of the evolution of the
universe but $s$ first decreases from positive value to negative
value and after that $s$ also increases to positive value.\\

\begin{figure}
\includegraphics[height=4in]{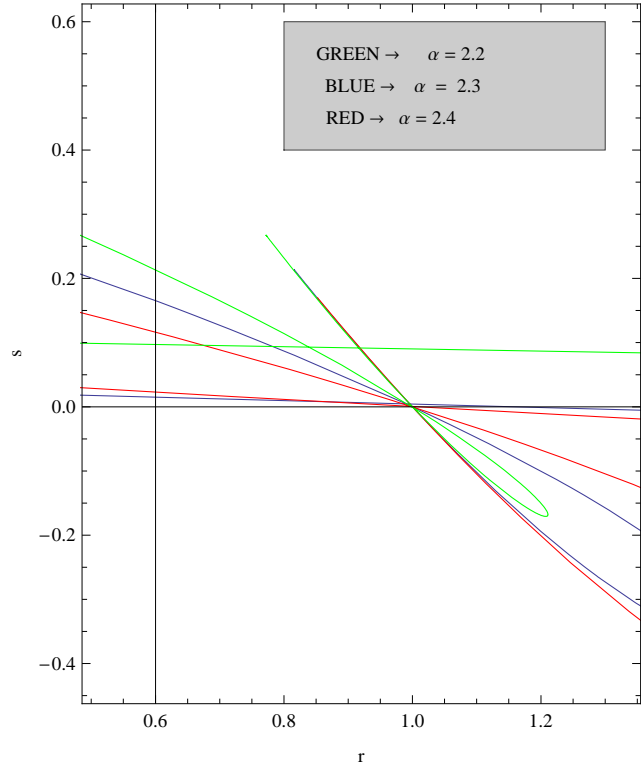}~~~~
\caption{The variation of $s$ against $r$ for logamediate Scenario
with $A =1$ and $\alpha=2.2, 2.3, 2.4$} \vspace{7mm}
\end{figure}

From (2) we get the total energy density of the universe,
\begin{equation}
\rho_{tot}=3H^2=\frac{3A^2\alpha^2(\ln t)^{2\alpha-2}}{t^2}
\end{equation}

\subsection{\normalsize\bf{Intermediate Scenario}}
Consider a particular form of Intermediate Scenario [25], where
the scale factor $a(t)$ of the Friedmann universe is described as,
\begin{equation}
a(t)=\exp (B t^\beta)
\end{equation}
 where $B\beta>0$, $B>0$ and $0<\beta<1$. Here the expansion of Universe is faster than Power-Law
form, where the scale factor is given as, $a(t) = t^n$, where
$n>1$ is a constant. Also, the expansion of the Universe is slower
for Standard de-Sitter Scenario where $\beta = 1$.
 The Hubble parameter $H=\frac{\dot{a}}{a}$ becomes,
\begin{equation}
H=B\beta t^{\beta-1}
\end{equation}
Hence from (15) we get,
\begin{equation}
\frac{\dot H}{H}=\frac{\beta-1}{t}
\end{equation}
and
\begin{equation}
\frac{\ddot H}{H}=\frac{(\beta-1)(\beta-2)}{t^2}
\end{equation}

Putting the value of $a(t)$ in the deceleration parameter
$q=-\frac{a {\ddot{a}}}{{\dot{a}}^2}$ we get,
\begin{equation}
q=-1-\frac{\beta-1}{B\beta t^\beta}
\end{equation}
where $a(t)$ is the scale factor. It has been seen that $q>-1$.
Fig.3 represents the variation of $H$ against $q$ for different values of $\beta$.\\

\begin{figure}
\includegraphics[height=2in]{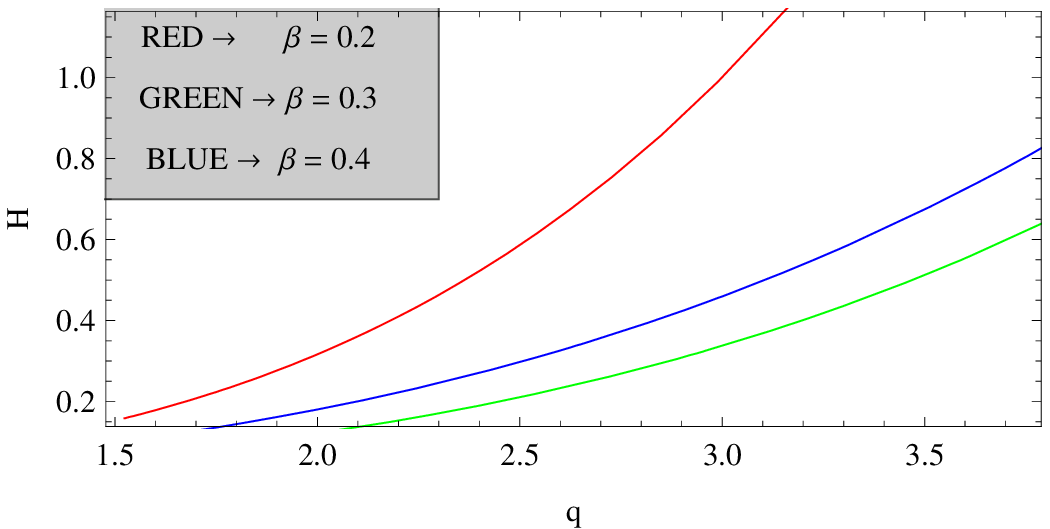}~~~~
\caption{The variation of $H$ against $q$ for intermediate
Scenario with $B = 1$ and $\beta=0.2,0.3,0.4$} \vspace{7mm}
\end{figure}

From (10), we get the expressions for statefinder parameters as
\begin{equation}
r=1+\frac{(\beta-1)(\beta-2)}{B^2 \beta^2}t^{-2\beta} +
\frac{\beta+1}{B\beta}t^{-\beta}
\end{equation}
and
\begin{equation}
s=-\frac{\frac{(\beta-1)(\beta-2)}{B\beta
t^\beta}+\beta+1}{3(\beta-1)+\frac{9B\beta t^\beta}{2}}
\end{equation}
Fig.4 represents the variation of $s$ against $r$ for different
values of $\beta$. We see that $r$ increases with increasing $s$.
At the evolution of the universe, $r$ and $s$ are both increase and
keep positive sign always.\\

From (2) we get the total energy density of the universe,
\begin{equation}
\rho_{tot}=3H^2=3B^2\beta^2t^{2\beta-2}
\end{equation}
\begin{figure}
\includegraphics[height=1.8in]{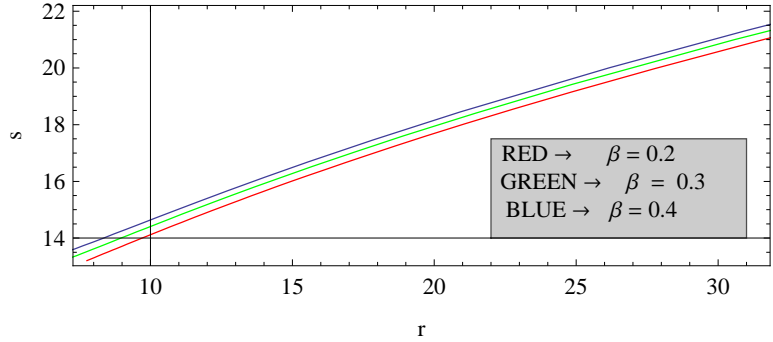}~~~~
\caption{The variation of $s$ against $r$ for intermediate
Scenario with $B = 1$ and $\beta=0.2, 0.3, 0.4$} \vspace{7mm}
\end{figure}

\section{\normalsize\bf{Mixture of Generalized Chaplygin
Gas with Tachyonic Field}}

The Lagrangian density for the tachyonic field is denoted as $\cal
L$, defined as [15],
\begin{equation}
{\cal {L}}=-V(\phi)\sqrt{1+g^{\mu \nu}~\partial{_{\mu}}\phi
\partial{_{\nu}} \phi}
\end{equation}
where $\phi$ is the tachyonic field and $V(\phi)$ is the tachyonic
potential. The homogeneous tachyon condensate of string theory in
a gravitational background is given by,
\begin{equation}
S=\int{\sqrt{-g}~ d ^{4} x \left[\frac{\cal R}{16 \pi G}+{\cal
L}\right]}
\end{equation}
where $\cal R$ is the Ricci Scalar. The energy-momentum tensor for
the tachyonic field is,
\begin{equation}
T_{\mu \nu}=-V(\phi)\sqrt{1+g^{\mu \nu} \partial _{\mu}\phi
\partial_{\nu} \phi}~g^{\mu \nu}+V(\phi) \frac{\partial _{\mu}\phi
\partial_{\nu} \phi}{\sqrt{1+g^{\mu \nu} \partial _{\mu}\phi \partial_{\nu} \phi}}
\end{equation}

where the velocity $u_{\mu}$ is given by,
\begin{equation}
u_{\mu}=-\frac{\partial_{\mu}\phi}{\sqrt{-g^{\mu \nu} \partial
_{\mu}\phi \partial_{\nu} \phi}}
\end{equation}
with $u^{\nu} u_{\nu}=-1$.\\

So the energy density $\rho_{t}$ and the pressure $p_{t}$ of the
tachyonic field $\phi$ become
\begin{equation}
\rho_{t}=\frac{V(\phi)}{\sqrt{1-{\dot{\phi}}^{2}}}~~~~~\text{and}~~~~~
p_{t}=-V(\phi) \sqrt{1-{\dot{\phi}}^{2}}
\end{equation}

Hence from (26) we get,

\begin{equation}
\phi=\int\sqrt{1+\frac{p_{t}}{\rho_{t}}}~dt
\end{equation}
and
\begin{equation}
V(\phi)=\sqrt{-p_{t} \rho_{t}}
\end{equation}
which represents pure Chaplygin gas if $V(\phi)$ is assumed as a constant
(i.e. $p_{t}$ and $\rho_{t}$ are inversely proportional).\\

In the class of scalar potentials, Barrow [25] has assumed
slow-roll inflation, $3H\dot{\phi}\approx -dV/d\phi$. Indeed, as
field rolls down the potential towards larger values, the
slow-roll approximation becomes increasingly more accurate. In the
Hamilton-Jacobi formalism, the slow-roll-parameters are defined as
[25],
\begin{equation}
\epsilon=2\left(\frac{H'}{H}\right)^2=\frac{2{\dot
H}^2}{H^2\dot{\phi}^{2}}
\end{equation}
and
\begin{equation}
\eta=\frac{2H''}{H}=\frac{2}{H}\left(\frac{\ddot
H}{\dot\phi^2}-\frac{\dot H \ddot\phi}{\dot\phi^3}\right)
\end{equation}
where DOT indicates differentiation w.r.t. $t$ and DASH indicates
differentiation w.r.t. $\phi$. Barrow [25] has shown that the
slow-roll parameter $\epsilon$ diverges when the field approaches
zero, has a minimum at the maximum of the potential, peaks at some
value $\phi_{\epsilon}$, and finally asymptotes to zero for large
values of the field. It has been shown that the peak occurs for
$\epsilon>1$, so that at the moment when inflation begins with
$\phi_{1}\equiv \phi(\epsilon=1)$.\\

For the accelerated expansion of the universe, we search a new
type of matter i.e., dark energy which violates the strong energy
condition. Pure Chaplygin Gas (PCG) is a particular type of dark
energy, which obeys an equation of state, $p=-C/\rho$ [8-12],
where $p$ and $\rho$ are the pressure and energy density of the
PCG respectively where $C$ is a positive constant. The PCG was
modified to generalized Chaplygin gas (MCG), which obeys an
equation of state, $p=-C/\rho^\gamma$ where $0 \leq\gamma\leq 1$.
The GCG is modified to {\bf Modified Chaplygin Gas} [13,14]
obeying an equation of state $p = A\rho - C/\rho^\gamma$ with
$0\leq\gamma\leq 1$, where $A,~ C$ are positive constants. This
equation of state shows radiation era at
one extreme and a $\Lambda$CDM model at the other extreme. \\

Let us consider the universe is filled with the mixture of
generalized Chaplygin Gas and tachyonic field. This generalized
Chaplygin Gas is considered a perfect fluid which follows the
adiabatic equation of state. The equation of Generalized Chaplygin
Gas is given by,
\begin{equation}
p_{c}=-C/{\rho}_{c}^{\gamma}~~,~~~ 0\le \gamma \le 1, C>0.
\end{equation}

If the energy density of the fluid is a function of volume only,
the temperature of the fluid remains zero at any pressure or
volume, violating the third law of thermodynamics.  The total
energy density and pressure are respectively given by,
\begin{equation}
\rho_{tot}=\rho_{c}+\rho_{t}
\end {equation}
\begin{equation} p_{tot}=p_{c}+p_{t}
\end {equation}
where $p_{c}$ and $\rho_{c}$ are the pressure and density of the
generalized Chaplygin gas respectively and $p_{t}$ and $\rho_{t}$
are the pressure and density of the Tachyonic field respectively.
Now we consider two possible states: (i) Without interaction and
(ii) With interaction.\\

\subsection{\normalsize\bf{Without Interaction}}
 The energy conservation equation is,
\begin{equation}
\dot{\rho}_{tot}+3\frac{\dot{a}}{a}(\rho_{tot}+p_{tot})=0
\end{equation}
Suppose two fluids do not interact with each other. Then the
above equation may be written as,
\begin{equation}
\dot{\rho}_{t}+3\frac{\dot{a}}{a}(\rho_{t}+p_{t})=0
\end{equation}
and
\begin{equation}
\dot{\rho}_{c}+3\frac{\dot{a}}{a}(\rho_{c}+p_{c})=0
\end{equation}

Now from equations (31) and (36), after eliminating $p_{c}$ we get
$\rho_{c}$ in terms of the scale factor,
\begin{equation}
\rho_{c}=\left[C+\rho_{0}
a^{-3(1+\gamma)}\right]^{\frac{1}{1+\gamma}}
\end{equation}
where $\rho_{0}$ is the integrating constant.\\\\

{\normalsize{\bf{Case I:}}}\\

In case of {\bf Logamediate Scenario} using (5), equation (37)
reduces to,
\begin{equation}
\rho_{c}=[C+\rho_{0}x_{1}]^{\frac{1}{(1+\gamma)}}
\end{equation}
where, $x_{1}=\exp (-3A(1+\gamma)(\ln t)^{\alpha})$.  Hence from
(13) and (38) the energy density of the tachyonic fluid becomes
\begin{equation}
\rho_{t}=\frac{3A^{2}\alpha^2 (\ln t)^{2\alpha-2}}{t^{2}}
-[C+\rho_{0}x_{1}]^{\frac{1}{(1+\gamma)}}
\end{equation}

Hence from (35) and (39) the pressure of the tachyonic fluid
becomes,
\begin{equation}
p_{t}=-\frac{3A^{2}\alpha^2 (\ln t)^{2\alpha-2}}{t^{2}}
+\frac{2A\alpha(\ln t -\alpha+1)(\ln
t)^{\alpha-2}}{t^2}+C[C+\rho_{0}x_{1}]^{\frac{-\gamma}{(1+\gamma)}}
\end{equation}

Solving the equations (27), (28), (39) and (40), the tachyonic
field and the tachyonic potential are obtained as,
\begin{equation}
\phi=\int\sqrt{\frac{\frac{2A\alpha(-\ln t + \alpha-1)(\ln
t)^{\alpha-2}}{t^{2}}+\rho_{0}x_{1}[C+\rho_{0}x_{1}]^\frac{-\gamma}{\gamma+1}}
{[C+\rho_{0}x_{1}]^\frac{1}{\gamma+1}-\frac{3A^2\alpha^2(\ln
t)^{2\alpha-2}}{t^2}}}~dt
\end{equation}
and
\begin{eqnarray*}
V(\phi)=\sqrt{\frac{3A^{2}\alpha^2 (\ln t)^{2\alpha-2}}{t^{2}}
-[C+\rho_{0}x_{1}]^{\frac{1}{(1+\gamma)}}}~~\times~~~~~~~~~~~~~~~~~~~~~~~~~~
\end{eqnarray*}
\begin{equation}
\sqrt{\frac{3A^{2}\alpha^2 (\ln t)^{2\alpha-2}}{t^{2}}
-\frac{2A\alpha(\ln t -\alpha+1)(\ln
t)^{\alpha-2}}{t^2}-C[C+\rho_{0}x_{1}]^{\frac{-\gamma}{(1+\gamma)}}}
\end{equation}

Fig.5 ~represents the variation of $V$ against $\phi$ for
different values of $\alpha$. In this case, the potential
always decreases with the tachyonic field $\phi$.\\
\begin{figure}
\includegraphics[height=1.5in]{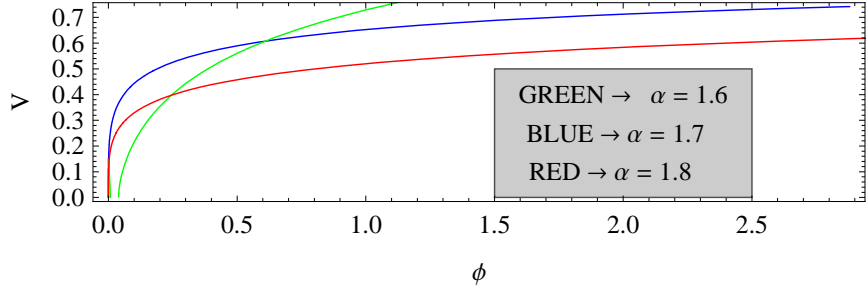}~~~~
\caption{The variation of $V$ against $\phi$ from equations (41)
and (42) for $A = C = 1, \rho_{0} = 5, \gamma=.5 $   and
$\alpha=1.6,1.7,1.8$} \vspace{7mm}
\end{figure}

From (7), (8), (29), (30) and (41) we get the slow-roll
parameters,
\begin{equation}
\epsilon=2\left(\frac{\alpha-1-\ln t}{t\ln t}\right)^2\times
\frac{[C+\rho_{0}x_{1}]^\frac{1}{\gamma+1}-\frac{3A^2\alpha^2(\ln
t)^{2\alpha-2}}{t^2}}{\frac{2A\alpha(-\ln t + \alpha-1)(\ln
t)^{\alpha-2}}{t^{2}}+\rho_{0}x_{1}[C+\rho_{0}x_{1}]^\frac{-\gamma}{\gamma+1}}
\end{equation}
and
\begin{eqnarray*}
\eta=2\times\frac{[C+\rho_{0}x_{1}]^\frac{1}{\gamma+1}-\frac{3A^2\alpha^2(\ln
t)^{2\alpha-2}}{t^2}}{\frac{2A\alpha(-\ln t + \alpha-1)(\ln
t)^{\alpha-2}}{t^{2}}+\rho_{0}x_{1}[C+\rho_{0}x_{1}]^\frac{-\gamma}{\gamma+1}}
\times\left(\frac{2(\ln t)^2-3(\alpha-1)\ln
t+(\alpha-1)(\alpha-2)}{t^2 (\ln t)^2}\right)-
\end{eqnarray*}
\begin{equation}
\left(\frac{\alpha-1-\ln t}{t\ln t}\right)
\left(\frac{[C+\rho_{0}x_{1}]^\frac{1}{\gamma+1}-\frac{3A^2\alpha^2(\ln
t)^{2\alpha-2}}{t^2}}{\frac{2A\alpha(-\ln t + \alpha-1)(\ln
t)^{\alpha-2}}{t^{2}}+\rho_{0}x_{1}[C+\rho_{0}x_{1}]^\frac{-\gamma}{\gamma+1}}\right)^2
\frac{\partial}{\partial{t}}\left[\frac{\frac{2A\alpha(-\ln t +
\alpha-1)(\ln
t)^{\alpha-2}}{t^{2}}+\rho_{0}x_{1}[C+\rho_{0}x_{1}]^\frac{-\gamma}{\gamma+1}}{[C+\rho_{0}x_{1}]^\frac{1}{\gamma+1}-\frac{3A^2\alpha^2(\ln
t)^{2\alpha-2}}{t^2}}\right]
\end{equation}

From the above equations, we see that $\eta$ can not be expressed
explicitly in terms of $\epsilon$. So we draw the graph of $\eta$
against $\epsilon$ in Fig.6 for different values of $\alpha$. \\

\begin{figure}
\includegraphics[height=3in]{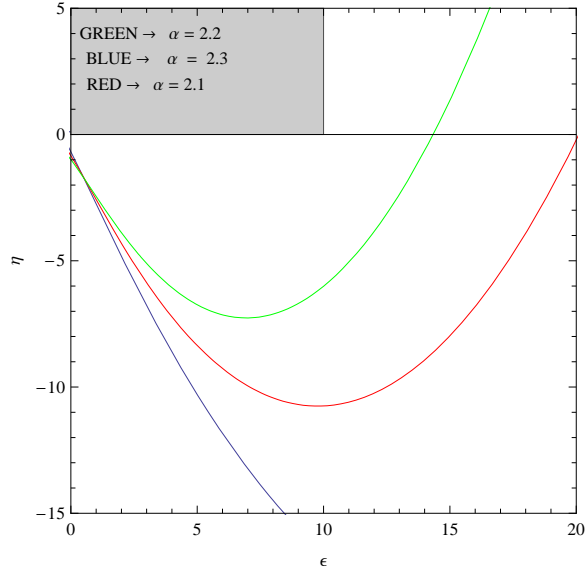}~~~~
\caption{The variation of $\eta$ against $\epsilon$ from equations
(43) and (44) for $A = C = 1, \rho_{0}=5,\gamma=.5$ and
$\alpha=2.1,2.2,2.3$} \vspace{7mm}
\end{figure}

{\normalsize{\bf{Case II:}}}\\

In case of {\bf{Intermediate Scenario}}, using (14), equation (37)
reduces to,
\begin{equation}
\rho_{c}=[C+\rho_{0}x_{2}]^{\frac{1}{(1+\gamma)}}
\end{equation}
where, $x_{2}=\exp (-3B(1+\gamma)t^\beta)$. \\

Hence from (21), (33), (35) and (45) we get the energy density and
the pressure of the tachyonic fluid is,
\begin{equation}
\rho_{t}=3B^2\beta^2
t^{2\beta-2}-[C+\rho_{0}x_{2}]^{\frac{1}{(1+\gamma)}}
\end{equation}
and
\begin{equation}
p_{t}=-3B^2\beta^2 t^{2\beta-2}-2B\beta(\beta-1)t^{\beta-2}
+C[C+\rho_{0}x_{2}]^{\frac{-\gamma}{(1+\gamma)}}
\end{equation}

Solving the equations (27), (28), (46) and (47), the tachyonic
field and the tachyonic potential are obtained as,
\begin{equation}
\phi=\int\sqrt{\frac{2B\beta
(\beta-1)t^{\beta-2}-C[C+\rho_{0}x_{2}]^{\frac{-\gamma}{(1+\gamma)}}}{[C+\rho_{0}x_{2}]
^{\frac{1}{(1+\gamma)}}-3B^2\beta^2 t^{2\beta-2}}}dt
\end{equation}
and
\begin{equation}
V(\phi)=\sqrt{3B^2\beta^2
t^{2\beta-2}-[C+\rho_{0}x_{2}]^{\frac{1}{(1+\gamma)}}}\times
\sqrt{3B^2\beta^2
t^{2\beta-2}+2B\beta(\beta-1)t^{\beta-2}-C[C+\rho_{0}x_{2}]^{\frac{-\gamma}{(1+\gamma)}}
}
\end{equation}

Fig.7 ~represents the variation of $V$ against $\phi$ for
different values of $\beta$. Here the potential $V$ is sharply
decreasing with the tachyonic field $\phi$.\\

\begin{figure}
\includegraphics[height=2.3in]{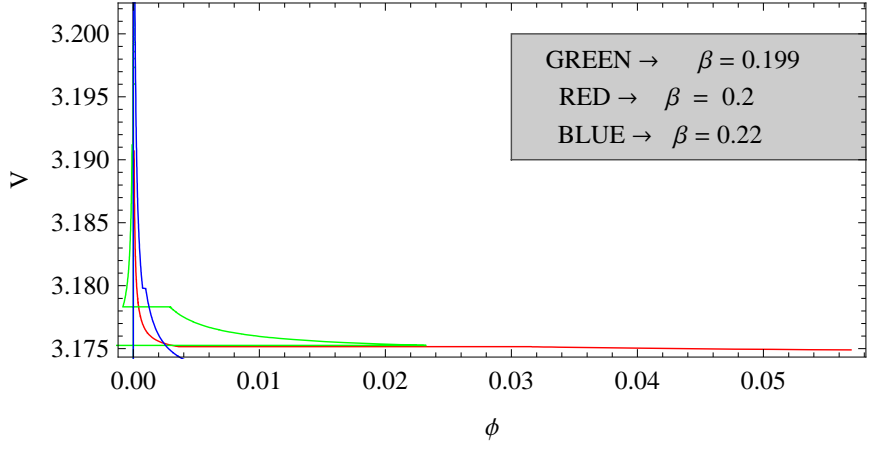}~~~~
\caption{The variation of $V$ against $\phi$ from (48) and (49)
for $C = 2, \rho_{0} = 5,\gamma=.5, B = 1$ and $\beta=0.199, 0.2,
0.22$} \vspace{7mm}
\end{figure}

From (16), (17), (29), (30) and (48) we get the slow-roll
parameter,
\begin{equation}
\epsilon=2\left(\frac{\beta-1}{t}\right)^2\times
\frac{[C+\rho_{0}x_{2}]^{\frac{1}{(1+\gamma)}}-3B^2\beta^2
t^{2\beta-2}}{2B\beta
(\beta-1)t^{\beta-2}-C[C+\rho_{0}x_{2}]^{\frac{-\gamma}{(1+\gamma)}}}
\end{equation}
and
\begin{eqnarray*}
\eta=\frac{2(\beta-1)(\beta-2)}{t^2}\times\frac{[C+\rho_{0}x_{2}]^{\frac{1}{(1+\gamma)}}-3B^2\beta^2
t^{2\beta-2}}{2B\beta
(\beta-1)t^{\beta-2}-C[C+\rho_{0}x_{2}]^{\frac{-\gamma}{(1+\gamma)}}}-
\end{eqnarray*}
\begin{equation}
\left(\frac{\beta-1}{t}\right)
\left(\frac{[C+\rho_{0}x_{2}]^{\frac{1}{(1+\gamma)}}-3B^2\beta^2
t^{2\beta-2}}{2B\beta
(\beta-1)t^{\beta-2}-C[C+\rho_{0}x_{2}]^{\frac{-\gamma}{(1+\gamma)}}}\right)^2
\frac{\partial}{\partial{t}}\left[\frac{2B\beta
(\beta-1)t^{\beta-2}-C[C+\rho_{0}x_{2}]^{\frac{-\gamma}{(1+\gamma)}}}{[C+\rho_{0}x_{2}]
^{\frac{1}{(1+\gamma)}}-3B^2\beta^2 t^{2\beta-2}}\right]
\end{equation}

Fig.8~ represents the variation of $\eta$ against $\epsilon$ for
different values of $\beta$. From this figure, it has been seen
that $\eta$ is sharply decreasing with increasing $\epsilon$.\\

\begin{figure}
\includegraphics[height=3.5in]{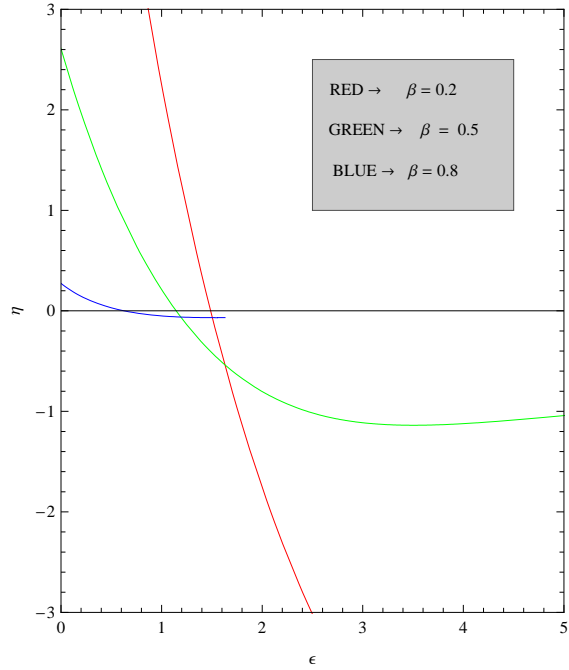}~~~~
\caption{The variation of $\eta$ against $\epsilon$ from (50) and
(51) for $C = 2,\gamma=.5, \rho_{0} = 5, B = 1$ and $\beta=0.2,
0.5, 0.8$} \vspace{7mm}
\end{figure}

\subsection{\normalsize\bf{With Interaction}}

Now we consider an interaction between the tachyonic fluid and GCG
by introducing an interaction term as a product of the Hubble
parameter and the energy density of the Chaplygin gas. Thus there
is an energy flow between the two fluids. \\

Now the equations of motion corresponding to the tachyonic field
and GCG are respectively,
\begin{equation}
\dot{\rho}_{t}+3\frac{\dot{a}}{a}(\rho_{t}+p_{t})=-3H\delta
\rho_{c}
\end{equation}
and
\begin{equation}
\dot{\rho}_{c}+3\frac{\dot{a}}{a}(\rho_{c}+p_{c})=3H\delta
\rho_{c}
\end{equation}
where $\delta$ is a coupling constant.\\

Solving equation (53) with the help of equations (14) and (31) we
get,
\begin{equation}
\rho_{c}=\left[\frac{C}{1-\delta}+\rho_{0}
a^{-3(1+\gamma)(1-\delta)}\right]^{\frac{1}{(1+\gamma)}}
\end{equation}

{\normalsize{\bf{Case I:}}}\\

In case of {\bf{Logamediate Scenario}}, we get the solutions:
\begin{equation}
\rho_{c}=\left[\frac{C}{1-\delta}+\rho_{0}x_{3}\right]^{\frac{1}{(1+\gamma)}}
\end{equation}
\begin{equation}
\rho_{t}=\frac{3A^{2}\alpha^2 (\ln
t)^{2\alpha-2}}{t^{2}}-\left[\frac{C}{1-\delta}+\rho_{0}
x_{3}\right]^{\frac{1}{(1+\gamma)}}
\end{equation}
where $x_3=\exp (-3A(1-\delta)(1+\gamma)(\ln t)^{\alpha})$. Hence,
\begin{equation}
p_{t}=-\frac{3A^{2}\alpha^2 (\ln t)^{2\alpha-2}}{t^{2}}
+\frac{2A\alpha(\ln t-\alpha+1)(\ln t)^{\alpha-2}}{t^2}
+C\left[\frac{C}{1-\delta}+\rho_{0}x_{3}\right]^{\frac{-\gamma}{(1+\gamma)}}
\end{equation}

Solving the equations, the tachyonic field is obtained as,
\begin{equation}
\phi=\int\sqrt{\frac{\frac{2A\alpha(\ln t-\alpha+1)(\ln
t)^{\alpha-2}}{t^2}-\left(\frac{C\delta}{1-\delta}+\rho_{0}x_{3}\right)\left(
\frac{C}{1-\delta}+\rho_{0}x_{3}\right)^\frac{-\gamma}{1+\gamma}}{\frac{3A^{2}\alpha^2
(\ln t)^{2\alpha-2}}{t^{2}}-\left[\frac{C}{1-\delta}+\rho_{0}
x_{3}\right]^{\frac{1}{(1+\gamma)}}}}~dt
\end{equation}
Also the potential will be of the form,\\
\begin{eqnarray*}
V(\phi)=\sqrt{\frac{3A^{2}\alpha^2 (\ln
t)^{2\alpha-2}}{t^{2}}-\left[\frac{C}{1-\delta}+\rho_{0}
x_{3}\right]^{\frac{1}{(1+\gamma)}}}~~\times
\end{eqnarray*}
\begin{equation}
\sqrt{\frac{3A^{2}\alpha^2 (\ln t)^{2\alpha-2}}{t^{2}}
-\frac{2A\alpha(\ln t-\alpha+1)(\ln t)^{\alpha-2}}{t^2}
-C\left[\frac{C}{1-\delta}+\rho_{0}x_{3}\right]^{\frac{-\gamma}{(1+\gamma)}}}
\end{equation}

In this case the potential starting from a large value and finally
tends to small value (Fig.9).\\\

\begin{figure}
\includegraphics[height=3in]{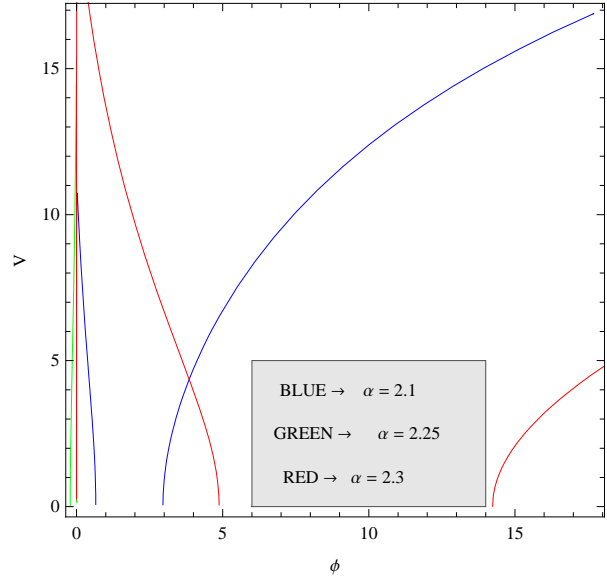}~~~~
\caption{The variation of $V$ against $\phi$ from (58) and (59)
for $A = C = 1, \rho_{0} = 5, \gamma=.5, \delta=.2$ and
$\alpha=2.1, 2.25, 2.3$} \vspace{7mm}
\end{figure}

\begin{figure}
\includegraphics[height=3in]{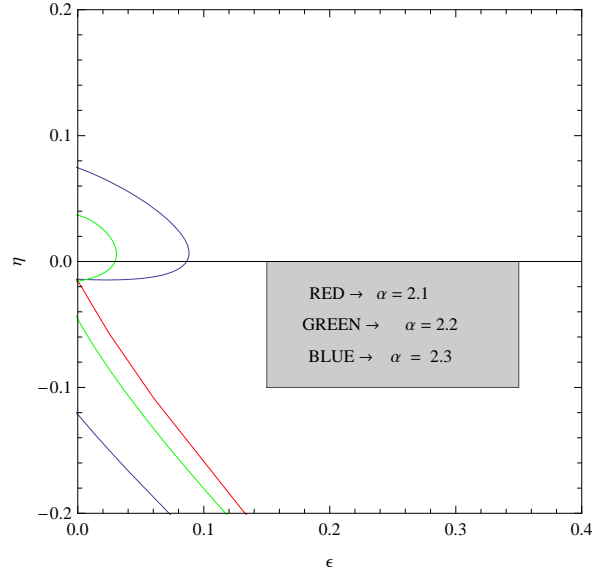}~~~~
\caption{The variation of $\eta$ against $\epsilon$ from (60) and
(61) for $A = C = 1, \rho_{0} = 5, \delta=.2,\gamma=.5$ and
$\alpha=2.1, 2.2, 2.3$} \vspace{7mm}
\end{figure}

The slow-roll parameters are obtained as

\begin{equation}
\epsilon=2\left(\frac{\alpha-1-\ln t}{t\ln t}\right)^2\times
\frac{\frac{3A^{2}\alpha^2 (\ln
t)^{2\alpha-2}}{t^{2}}-\left[\frac{C}{1-\delta}+\rho_{0}
x_{3}\right]^{\frac{1}{(1+\gamma)}}}{\frac{2A\alpha(\ln
t-\alpha+1)(\ln
t)^{\alpha-2}}{t^2}-\left(\frac{C\delta}{1-\delta}+\rho_{0}x_{3}\right)\left(
\frac{C}{1-\delta}+\rho_{0}x_{3}\right)^\frac{-\gamma}{1+\gamma}}
\end{equation}
and
\begin{eqnarray*}
\eta=2\times\frac{\frac{3A^{2}\alpha^2 (\ln
t)^{2\alpha-2}}{t^{2}}-\left[\frac{C}{1-\delta}+\rho_{0}
x_{3}\right]^{\frac{1}{(1+\gamma)}}}{\frac{2A\alpha(\ln
t-\alpha+1)(\ln
t)^{\alpha-2}}{t^2}-\left(\frac{C\delta}{1-\delta}+\rho_{0}x_{3}\right)\left(
\frac{C}{1-\delta}+\rho_{0}x_{3}\right)^\frac{-\gamma}{1+\gamma}}\times
\frac{2(\ln t)^2-3(\alpha-1)\ln t+(\alpha-1)(\alpha-2)}{t^2 (\ln
t)^2}-
\end{eqnarray*}
\begin{eqnarray*}
\left(\frac{\alpha-1-\ln t}{t\ln t}\right)\times
\frac{\frac{3A^{2}\alpha^2 (\ln
t)^{2\alpha-2}}{t^{2}}-\left[\frac{C}{1-\delta}+\rho_{0}
x_{3}\right]^{\frac{1}{(1+\gamma)}}}{\frac{2A\alpha(\ln
t-\alpha+1)(\ln
t)^{\alpha-2}}{t^2}-\left(\frac{C\delta}{1-\delta}+\rho_{0}x_{3}\right)\left(
\frac{C}{1-\delta}+\rho_{0}x_{3}\right)^\frac{-\gamma}{1+\gamma}}\times
\end{eqnarray*}
\begin{equation}
\frac{\partial}{\partial{t}}\left[\frac{\frac{2A\alpha(\ln
t-\alpha+1)(\ln
t)^{\alpha-2}}{t^2}-\left(\frac{C\delta}{1-\delta}+\rho_{0}x_{3}\right)\left(
\frac{C}{1-\delta}+\rho_{0}x_{3}\right)^\frac{-\gamma}{1+\gamma}}{\frac{3A^{2}\alpha^2
(\ln t)^{2\alpha-2}}{t^{2}}-\left[\frac{C}{1-\delta}+\rho_{0}
x_{3}\right]^{\frac{1}{(1+\gamma)}}}\right]
\end{equation}

From above expressions of $\epsilon$ and $\eta$, we see that
$\eta$ can not be expressed in terms of $\epsilon$. So we have
drawn the graph of $\eta$ against $\epsilon$ in Fig.10.\\

{\normalsize{\bf{Case II:}}}\\

In case of Intermediate Scenario, using (1), equation (36) reduces
to,
$$
\rho_{c}=\left[\frac{C}{1-\delta}+\rho_{0}x_{4}\right]^{\frac{1}{(1+\gamma)}}
$$
where $x_{4}=\exp (-3B(1-\delta)(1+\gamma)t^\beta)$. Hence the
energy density of the tachyonic fluid is,
\begin{equation}
\rho_{t}=3B^2\beta^2
t^{2\beta-2}-\left[\frac{C}{1-\delta}+\rho_{0}x_{4}\right]^{\frac{1}{(1+\gamma)}}
\end{equation}

Hence the pressure of the tachyonic fluid is,
\begin{equation}
p_{t}=-3B^2\beta^2
t^{2\beta-2}-2B\beta(\beta-1)t^{\beta-2}-\rho_{0}(1-\delta)(1+\gamma)
x_{4}\left[\frac{C}{1-\delta}+\rho_{0}x_{4}\right]^{\frac{-\gamma}{(1+\gamma)}}
\end{equation}

Solving the equations the tachyonic field and the tachyonic
potential are obtained as,
\begin{equation}
\phi=\int\sqrt{\frac{2B\beta(\beta-1)t^{\beta-2}-\left[\frac{C}{1-\delta}+\rho_{0}x_{4}\right]^{\frac{1}{(1+\gamma)}}
+\rho_{0}(1-\delta)(1+\gamma)x_{4}\left[\frac{C}{1-\delta}+\rho_{0}x_{4}\right]^{\frac{-\gamma}{(1+\gamma)}}}
{\left[\frac{C}{1-\delta}+\rho_{0}x_{4}\right]
^{\frac{1}{(1+\gamma)}}-3B^2\beta^2 t^{2\beta-2}}}~dt
\end{equation}

and
\begin{eqnarray*}
V(\phi)=\sqrt{3B^2\beta^2
t^{2\beta-2}-\left[\frac{C}{1-\delta}+\rho_{0}x_{4}\right]^{\frac{1}{(1+\gamma)}}}~~\times
\end{eqnarray*}
\begin{equation}
\sqrt{3B^2\beta^2
t^{2\beta-2}+2B\beta(\beta-1)t^{\beta-2}+\rho_{0}(1-\delta)(1+\gamma)
x_{4}\left[\frac{C}{1-\delta}+\rho_{0}x_{4}\right]^{\frac{-\gamma}{(1+\gamma)}}}
\end{equation}

\begin{figure}
\includegraphics[height=3in]{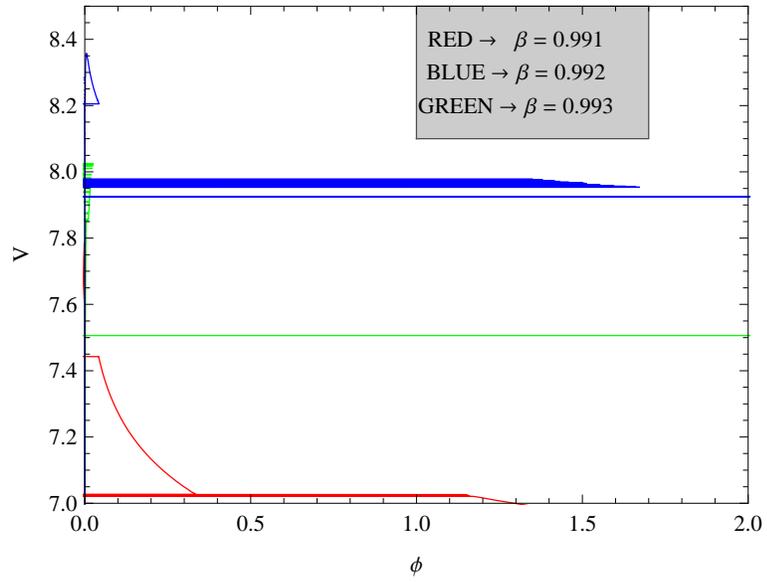}~~~~
\caption{The variation of $V$ against $\phi$ from (64) and (65)
for $B = 1,C = 2, \rho_{0} = 5, \gamma=\delta=.5$ and
$\beta=0.991, 0.992, 0.993$} \vspace{7mm}
\end{figure}

\begin{figure}
\includegraphics[height=4.5in]{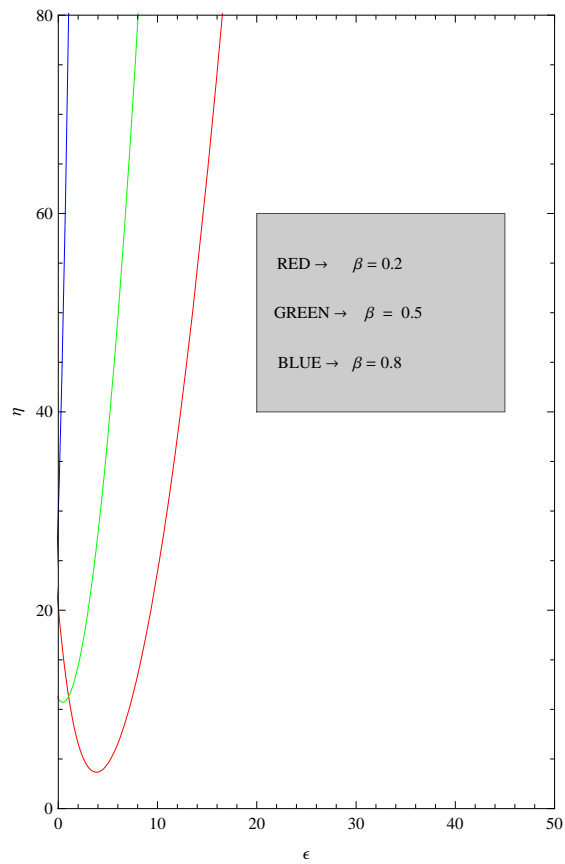}~~~~
\caption{The variation of $\eta$ against $\epsilon$ from (66) and
(67) for $B = 1,C = 2, \rho_{0} = 5, \gamma=\delta=.5$ and
$\beta=0.2, 0.5, 0.8$} \vspace{7mm}
\end{figure}

Fig.11 ~represents the variation of $V$ against $\phi$ for
different values of $\beta$. Here the potential $V$
decreases with the tachyonic field $\phi$.\\

The slow-roll parameters will be,
\begin{equation}
\epsilon=2\left(\frac{\beta-1}{t}\right)^2\times
\frac{\left[\frac{C}{1-\delta}+\rho_{0}x_{4}\right]^{\frac{1}{(1+\gamma)}}-3B^2\beta^2
t^{2\beta-2}}{2B\beta(\beta-1)t^{\beta-2}-\left[\frac{C}{1-\delta}+\rho_{0}x_{4}\right]^{\frac{1}{(1+\gamma)}}
+\rho_{0}(1-\delta)(1+\gamma)x_{4}\left[\frac{C}{1-\delta}+\rho_{0}x_{4}\right]^{\frac{-\gamma}{(1+\gamma)}}}
\end{equation}
and
\begin{eqnarray*}
\eta=\frac{2(\beta-1)(\beta-2)}{t^2}\times\frac{\left[\frac{C}{1-\delta}+\rho_{0}x_{4}\right]^{\frac{1}{(1+\gamma)}}-3B^2\beta^2
t^{2\beta-2}}{2B\beta(\beta-1)t^{\beta-2}-\left[\frac{C}{1-\delta}+\rho_{0}x_{4}\right]^{\frac{1}{(1+\gamma)}}
+\rho_{0}(1-\delta)(1+\gamma)
x_{4}\left[\frac{C}{1-\delta}+\rho_{0}x_{4}\right]^{\frac{-\gamma}{(1+\gamma)}}}
\end{eqnarray*}
\begin{eqnarray*}
-\left(\frac{\beta-1}{t}\right)\times\left(\frac{\left[\frac{C}{1-\delta}+\rho_{0}x_{4}\right]^{\frac{1}{(1+\gamma)}}-3B^2\beta^2
t^{2\beta-2}}{2B\beta(\beta-1)t^{\beta-2}-\left[\frac{C}{1-\delta}+\rho_{0}x_{4}\right]^{\frac{1}{(1+\gamma)}}
+\rho_{0}(1-\delta)(1+\gamma)x_{4}\left[\frac{C}{1-\delta}+\rho_{0}x_{4}\right]^{\frac{-\gamma}{(1+\gamma)}}}\right)^2\times
\end{eqnarray*}
\begin{equation}
\frac{\partial}{\partial{t}}\left[\frac{2B\beta(\beta-1)t^{\beta-2}-
\left[\frac{C}{1-\delta}+\rho_{0}x_{4}\right]^{\frac{1}{(1+\gamma)}}
+\rho_{0}(1-\delta)(1+\gamma)
x_{4}\left[\frac{C}{1-\delta}+\rho_{0}x_{4}\right]^{\frac{-\gamma}{(1+\gamma)}}}{\left[\frac{C}{1-\delta}+
\rho_{0}x_{4}\right]^{\frac{1}{(1+\gamma)}}-3B^2\beta^2
t^{2\beta-2}}\right]
\end{equation}

From fig.12, it has been seen that $\eta$ first decreases then
increases with $\epsilon$.\\

\section{\normalsize\bf{Mixture of Barotropic Fluid with Tachyonic
Field}}

A barotropic fluid is defined as that state of a fluid for which
is a function of only the pressure. The condition of barotropy of
a fluid represents another rather idealized state. However, in
this case the situation is closer to reality since compressibility
is allowed for. The term ``barotropic'' infers ``turning with (or
in the same manner as) the isobars'', referring to the isopycnals.
The name is a lucid one since it is obvious that if depends only
on $p$ then the isopycnal surfaces must always be parallel to the
isobaric surfaces, hence any change in inclination of the latter
brings about an identical change in orientation of the isopycnal
surfaces. The spacing of the isobaric surfaces with respect to
under quasistatic conditions depends only on $p$ for a barotropic
fluid. Furthermore, since is increased with increasing pressure
for a compressible fluid it is apparent that the spacing of
isobaric surfaces (for equal increments of $p$) relative to will
decrease with increasing $p$.\\

A fluid under conditions of perfect hydrostatic balance would
assume a barotropic state for which the pressure gradient can be
represented as a function of $p$ alone. However, this is a very
special case of barotropy where the isobaric surfaces are level.
Now we consider a two fluid model consisting of tachyonic field
and barotropic fluid. The EOS of the barotropic fluid is given by,
\begin{equation}
p_{b}=\omega_{b} \rho_{b}
\end{equation}

where $p_{b}$ and $\rho_{b}$ are the pressure and energy density
of the barotropic fluid. Hence the total energy density and
pressure are respectively given by,
\begin{equation}
\rho_{tot}=\rho_{b}+\rho_{t}
\end {equation}
and
\begin{equation}
p_{tot}=p_{b}+p_{t}
\end {equation}

\subsection{\normalsize\bf{Without Interaction}}

First we consider that the two fluids do not interact with each
other so that they are conserved separately. Therefore, the
conservation equation (34) reduces to,
\begin{equation}
\dot{\rho}_{t}+3\frac{\dot{a}}{a}(\rho_{t}+p_{t})=0
\end{equation}
and
\begin{equation}
\dot{\rho}_{b}+3\frac{\dot{a}}{a}(\rho_{b}+p_{b})=0
\end{equation}

Equation (72) together with equation (68) give,
\begin{equation}
\rho_{b}=\rho_{0}~a^{-3(1+\omega_{b})}
\end{equation}

{\normalsize{\bf{Case I:}}}\\

In case of Logamediate Scenario, using (5), equation (73) reduces
to,
\begin{equation}
\rho_{b}=\rho_{0}\exp (-3A(1+\omega_{b})(\ln t)^{\alpha})
\end{equation}

\begin{figure}
\includegraphics[height=3.5in]{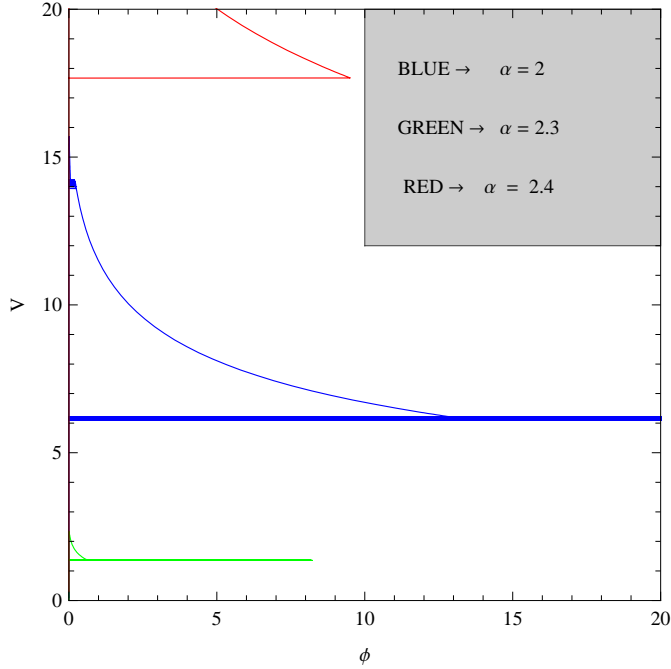}~~~~
\caption{The variation of $V$ against $\phi$ from (77) and (78)
for $A = 1, \rho_{0} = 5,\omega=.2$ and $\alpha=2, 2.3, 2.4$}
\vspace{7mm}
\end{figure}

Hence the energy density of the tachyonic fluid is,
\begin{equation}
\rho_{t}=\frac{3A^{2}\alpha^2 (\ln t)^{2\alpha-2}}{t^{2}}
-\rho_{0}x_{5}
\end{equation}
where, $x_{5}=\exp (-3A(1+\omega_{b})(\ln t)^{\alpha})$. Hence the
pressure of the tachyonic fluid is,
\begin{equation}
p_{t}=-\frac{3A^{2}\alpha^2 (\ln t)^{2\alpha-2}}{t^{2}}
+\frac{2A\alpha(\ln t -\alpha+1)(\ln
t)^{\alpha-2}}{t^2}-\rho_{0}\omega_{b}x_{5}
\end{equation}

Solving the equations the tachyonic field and the tachyonic
potential are obtained as,
\begin{equation}
\phi=\int\sqrt{\frac{\frac{2A\alpha(\ln t-\alpha+1)(\ln
t)^{\alpha-2}}{t^2}
-\rho_{0}(1+\omega_{b})x_{5}}{\frac{3A^{2}\alpha^2 (\ln
t)^{2\alpha-2}}{t^{2}} -\rho_{0}x_{5}}}dt
\end{equation}

and
\begin{equation}
V(\phi)=\sqrt{\frac{3A^{2}\alpha^2 (\ln t)^{2\alpha-2}}{t^{2}}
-\rho_{0}x_{5}}\times \sqrt{\frac{3A^{2}\alpha^2 (\ln
t)^{2\alpha-2}}{t^{2}} -\frac{2A\alpha(\ln t -\alpha+1)(\ln
t)^{\alpha-2}}{t^2}+\rho_{0}\omega_{b}x_{5}}
\end{equation}

From above equations, it has been seen that $V$ can not be
expressed in terms of $\phi$ explicitly. Fig. 13 shows the
variation of $V$ in terms of $\phi$.\\

The slow-roll parameters are obtained as,

\begin{equation}
\epsilon=2\left(\frac{\alpha-1-\ln t}{t\ln t}\right)^2\times
\frac{\frac{3A^{2}\alpha^2 (\ln t)^{2\alpha-2}}{t^{2}}
-\rho_{0}x_{5}}{\frac{2A\alpha(\ln t-\alpha+1)(\ln
t)^{\alpha-2}}{t^2} -\rho_{0}(1+\omega_{b})x_{5}}
\end{equation}
and
\begin{eqnarray*}
\eta=2\times\frac{\frac{3A^{2}\alpha^2 (\ln t)^{2\alpha-2}}{t^{2}}
-\rho_{0}x_{5}}{\frac{2A\alpha(\ln t-\alpha+1)(\ln
t)^{\alpha-2}}{t^2} -\rho_{0}(1+\omega_{b})x_{5}} \times
\left(\frac{2(\ln t)^2-3(\alpha-1)\ln t+(\alpha-1)(\alpha-2)}{t^2
(\ln t)^2}\right)
\end{eqnarray*}
\begin{equation}
-\left(\frac{\alpha-1-\ln t}{t\ln t}\right)\times
\left(\frac{\frac{3A^{2}\alpha^2 (\ln t)^{2\alpha-2}}{t^{2}}
-x_{5}}{\frac{2A\alpha(\ln t-\alpha+1)(\ln t)^{\alpha-2}}{t^2}
-\rho_{0}(1+\omega_{b})x_{5}}\right)^2
 \times\frac{\partial}{\partial{t}}\left[\frac{\frac{2A\alpha(\ln
t-\alpha+1)(\ln t)^{\alpha-2}}{t^2}
-\rho_{0}(1+\omega_{b})x_{5}}{\frac{3A^{2}\alpha^2 (\ln
t)^{2\alpha-2}}{t^{2}} -\rho_{0}x_{5}}\right]
\end{equation}

From complicated forms of $\eta$ and $\epsilon$, it has been seen
that $\eta$ can not be expressed in terms of $\epsilon$
explicitly. So we have shown the graph of $\eta$ with $\epsilon$
in fig. 14. \\

\begin{figure}
\includegraphics[height=3.5in]{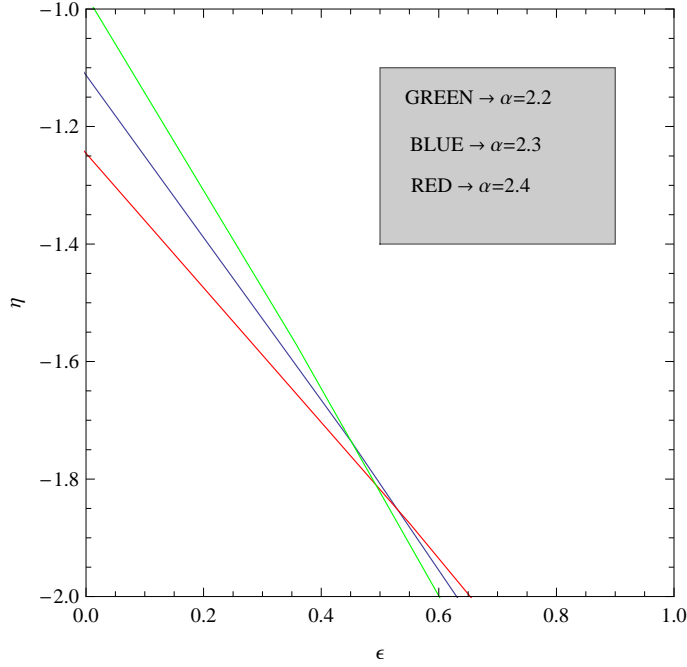}~~~~
\caption{The variation of $\eta$ against $\epsilon$ from (79) and
(80) for $A = 1,\omega_{b}=.2,\rho_{0}=5 $ and $\alpha=2.2, 2.3,
2.4$} \vspace{7mm}
\end{figure}

{\normalsize{\bf{Case II:}}}\\

In case of Intermediate Scenario, using (14), equation (73)
reduces to,
\begin{equation}
\rho_{b}=\rho_{0}\exp (-3B(1+\omega_{b})t^\beta)
\end{equation}

Hence the energy density of the tachyonic fluid is,
\begin{equation}
\rho_{t}=3B^2\beta^2 t^{2\beta-2}-\rho_{0}x_{6}
\end{equation}
where, $x_{6}=\exp(-3B(1+\omega_{b})t^\beta)$. Hence the pressure
of the tachyonic fluid is,
\begin{equation}
p_{t}=-3B^2\beta^2
t^{2\beta-2}-2B\beta(\beta-1)t^{\beta-2}-\rho_{0}\omega_{b}x_{6}
\end{equation}

Solving the equations the tachyonic field and the tachyonic
potential are obtained as,
\begin{equation}
\phi=\int\sqrt{\frac{2B\beta
(\beta-1)t^{\beta-2}+\rho_{0}(1+\omega_{b})x_{6}}{\rho_{0}x_{6}-3B^2\beta^2
t^{2\beta-2}}}dt
\end{equation}
and
\begin{equation}
V(\phi)=\sqrt{3B^2\beta^2 t^{2\beta-2}-\rho_{0}x_{6}}\times
\sqrt{3B^2\beta^2
t^{2\beta-2}+2B\beta(\beta-1)t^{\beta-2}+\rho_{0}\omega_{b}x_{6}}
\end{equation}

Like the mixture of tachyonic fluid with barotropic fluid in this
case also the potential $V$ starting from a low value increases
largely and then decreases to $0$ with time as shown in figure 15.\\

\begin{figure}
\includegraphics[height=3.5in]{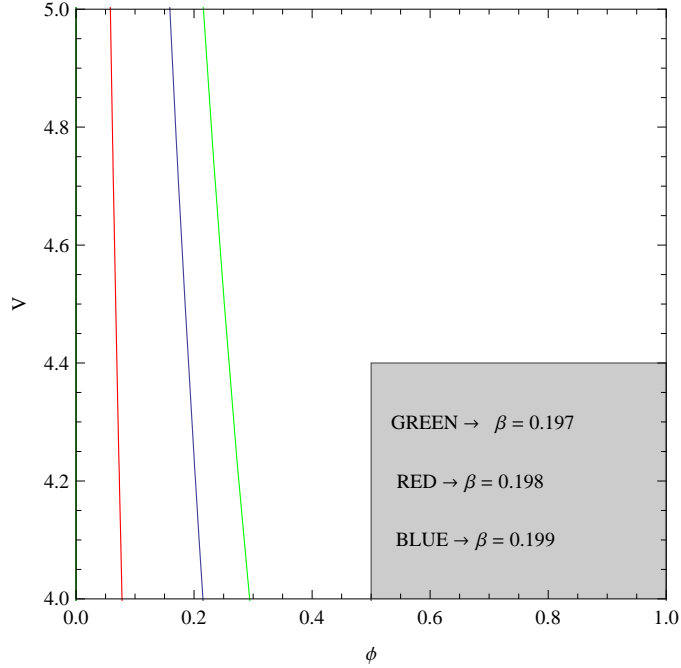}~~~~
\caption{The variation of $V$ against $\phi$ from (84) and (85)
for $B = 1,\omega_{b}=\frac{1}{3}, \rho_{0}=5$ and $\beta=0.197,
0.198, 0.199$} \vspace{7mm}
\end{figure}

The slow-roll parameters will be,
\begin{equation}
\epsilon=2\left(\frac{\beta-1}{t}\right)^2\times
\frac{\rho_{0}x_{6}-3B^2\beta^2 t^{2\beta-2}}{2B\beta
(\beta-1)t^{\beta-2}+\rho_{0}(1+\omega_{b})x_{6}}
\end{equation}
and
\begin{eqnarray*}
\eta=\frac{2(\beta-1)(\beta-2)}{t^2}\times\frac{\rho_{0}x_{6}-3B^2\beta^2
t^{2\beta-2}}{2B\beta
(\beta-1)t^{\beta-2}+\rho_{0}(1+\omega_{b})x_{6}}-\left(\frac{\beta-1}{t}\right)\times
\end{eqnarray*}
\begin{equation}
\left(\frac{\rho_{0}x_{6}-3B^2\beta^2 t^{2\beta-2}}{2B\beta
(\beta-1)t^{\beta-2}+\rho_{0}(1+\omega_{b})x_{6}}\right)^2\times
\frac{\partial}{\partial{t}}\left[\frac{2B\beta
(\beta-1)t^{\beta-2}+\rho_{0}(1+\omega_{b})x_{6}}{\rho_{0}x_{6}-3B^2\beta^2
t^{2\beta-2}}\right]
\end{equation}

From Fig.16 it has been seen that $\eta$ always decreases with
$\epsilon$.\\

\begin{figure}
\includegraphics[height=3.5in]{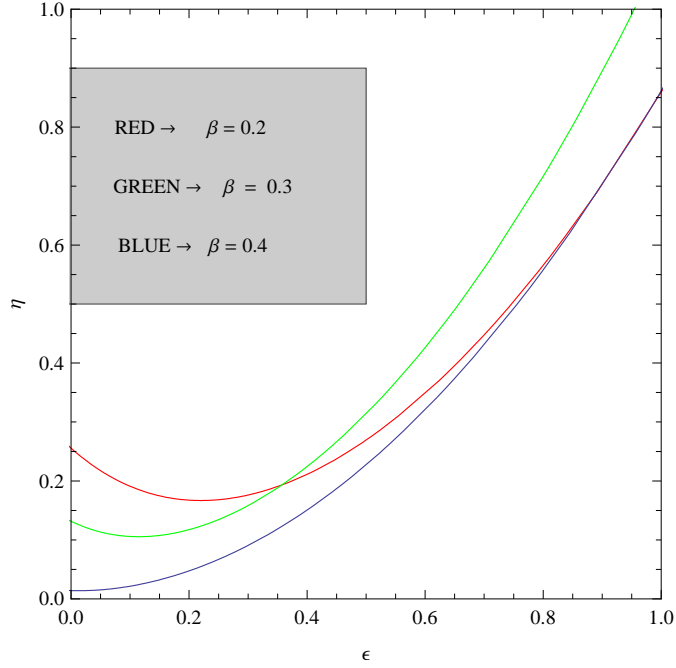}~~~~
\caption{The variation of $\eta$ against $\epsilon$ from (86) and
(87) for $B = 1,\omega_{b}=1/3,\rho_{0}=5 $ and $\beta=0.2, 0.3,
0.4$} \vspace{7mm}
\end{figure}

\subsection{\normalsize\bf{With Interaction}}

Now we consider an interaction between the tachyonic field and
the barotropic fluid by introducing a phenomenological coupling
function which is a product of the Hubble parameter and the
energy density of the barotropic fluid. Thus there is an energy
flow between the two fluids. \\

Now the equations of motion corresponding to the tachyonic field
and the barotropic fluid are respectively,
\begin{equation}
\dot{\rho}_{t}+3\frac{\dot{a}}{a}(\rho_{t}+p_{t})=-3H\delta
\rho_{b}
\end{equation}
and
\begin{equation}
\dot{\rho}_{b}+3\frac{\dot{a}}{a}(\rho_{b}+p_{b})=3H\delta
\rho_{b}
\end{equation}
 where $\delta$ is a coupling constant.\\

Solving equation (89) with the help of equation (68), we get,
\begin{equation}
\rho_{b}=\rho_{0}~a^{-3(1+\omega_{b}-\delta)}
\end{equation}

{\normalsize{\bf{Case I:}}}\\

In case of Logamediate Scenario, we obtain
\begin{equation}
\rho_{t}=\frac{3A^{2}\alpha^2 (\ln
t)^{2\alpha-2}}{t^{2}}-\rho_{0}x_{7}
\end{equation}
where, $x_{7}=\exp (-3A(1+\omega_{b}-\delta)(\ln t)^{\alpha})$.
Hence,
\begin{equation}
p_{t}=-\frac{3A^{2}\alpha^2 (\ln t)^{2\alpha-2}}{t^{2}}
+\frac{2A\alpha(\ln t-\alpha+1)(\ln
t)^{\alpha-2}}{t^2}-\rho_{0}\omega_{b}x_{7}
\end{equation}

Solving the equations the tachyonic field is obtained as,
\begin{equation}
\phi=\int\sqrt{\frac{\frac{2A\alpha(\ln t-\alpha+1)(\ln
t)^{\alpha-2}}{t^2}-\rho_{0}(1+\omega_{b})x_{7}}{\frac{3A^{2}\alpha^2
(\ln t)^{2\alpha-2}}{t^{2}}-\rho_{0} x_{7}}}dt
\end{equation}
Also the potential will be of the form,\\
\begin{equation}
V(\phi)=\sqrt{\frac{3A^{2}\alpha^2 (\ln
t)^{2\alpha-2}}{t^{2}}-\rho_{0} x_{7}}\times
\sqrt{\frac{3A^{2}\alpha^2 (\ln
t)^{2\alpha-2}}{t^{2}}-\frac{2A\alpha(\ln t-\alpha+1)(\ln
t)^{\alpha-2}}{t^2}+\rho_{0}\omega_{b}x_{7}}
\end{equation}

\begin{figure}
\includegraphics[height=2in]{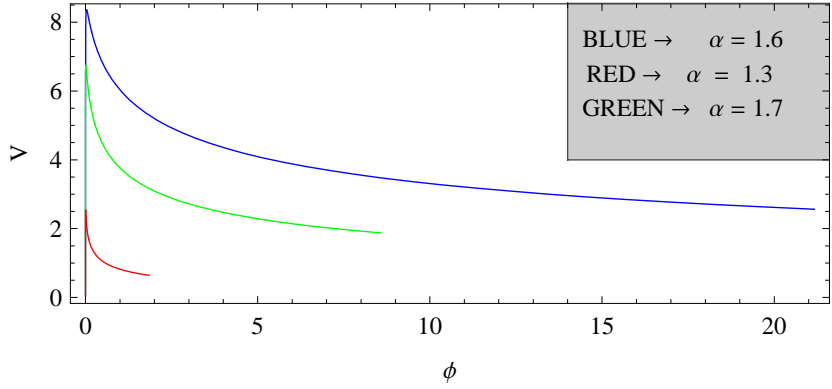}~~~~
\caption{The variation of $V$ against $\phi$ from (93) and (94)
for $A = 1,\omega_{b}=.2,\rho_{0}=5,\delta=.5$ and
$\alpha=1.3,1.6,1.7$} \vspace{7mm}
\end{figure}

The slow-roll parameters are obtained as

\begin{equation}
\epsilon=2\left(\frac{\alpha-1-\ln t}{t\ln t}\right)^2\times
\frac{\frac{3A^{2}\alpha^2 (\ln t)^{2\alpha-2}}{t^{2}}-\rho_{0}
x_{7}}{\frac{2A\alpha(\ln t-\alpha+1)(\ln
t)^{\alpha-2}}{t^2}-\rho_{0}(1+\omega_{b})x_{7}}
\end{equation}
and
\begin{eqnarray*}
\eta=2\times\frac{\frac{3A^{2}\alpha^2 (\ln
t)^{2\alpha-2}}{t^{2}}-\rho_{0}x_{7}}{\frac{2A\alpha(\ln
t-\alpha+1)(\ln t)^{\alpha-2}}{t^2}-\rho_{0}(1+\omega_{b})x_{7}}
\times\frac{2(\ln t)^2-3(\alpha-1)\ln t+(\alpha-1)(\alpha-2)}{t^2
(\ln t)^2}
\end{eqnarray*}
\begin{equation}
-\left(\frac{\alpha-1-\ln t}{t\ln t}\right)\times
\left(\frac{\frac{3A^{2}\alpha^2 (\ln
t)^{2\alpha-2}}{t^{2}}-\rho_{0} x_{7}}{\frac{2A\alpha(\ln
t-\alpha+1)(\ln
t)^{\alpha-2}}{t^2}-\rho_{0}(1+\omega_{b})x_{7}}\right)^2
\times\frac{\partial}{\partial{t}}\left[\frac{\frac{2A\alpha(\ln
t-\alpha+1)(\ln
t)^{\alpha-2}}{t^2}-\rho_{0}(1+\omega_{b})x_{7}}{\frac{3A^{2}\alpha^2
(\ln t)^{2\alpha-2}}{t^{2}}-\rho_{0} x_{7}}\right]
\end{equation}

\begin{figure}
\includegraphics[height=3.5in]{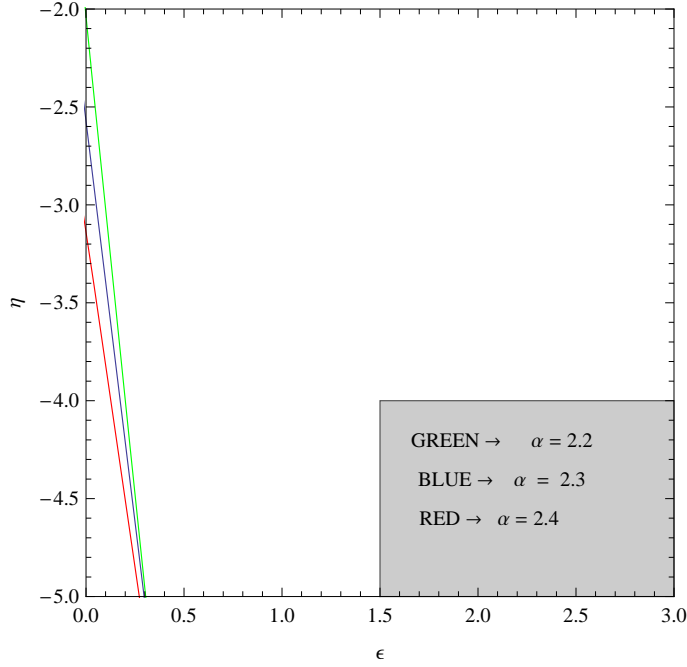}~~~~
\caption{The variation of $\eta$ against $\epsilon$ from (95) and
(96) for $A = 1,\omega_{b}=.2,\rho_{0}=5,\delta=.5$ and
$\alpha=2.2,2.3,2.4$} \vspace{7mm}
\end{figure}

Fig. 17 shows the variation of $V$ against $\phi$. It has been
seen that $V$ decreases as $\phi$ increases. Also from fig. 18, it
has been seen that $\eta$ decreases as $\epsilon$ increases.\\

{\normalsize{\bf{Case II:}}}\\

In case of Intermediate Scenario, using (14), equation (90)
reduces to,
\begin{equation}
\rho_{b}=\rho_{0}\exp(-3B(1+\omega_{b}-\delta)t^\beta)
\end{equation}

Hence the energy density of the tachyonic fluid is,
\begin{equation}
\rho_{t}=3B^2\beta^2 t^{2\beta-2}-\rho_{0}x_{8}
\end{equation}
where, $x_{8}=\exp(-3B(1+\omega_{b}-\delta)t^\beta)$. Hence the
pressure of the tachyonic fluid is,
\begin{equation}
p_{t}=-3B^2\beta^2
t^{2\beta-2}-2B\beta(\beta-1)t^{\beta-2}-\rho_{0}\omega_{b}x_{8}
\end{equation}

Solving the equations the tachyonic field and the tachyonic
potential are obtained as,
\begin{equation}
\phi=\int\sqrt{\frac{2B\beta(\beta-1)t^{\beta-2}+\rho_{0}\omega_{b}x_{8}}
{\rho_{0}x_{8}-3B^2\beta^2 t^{2\beta-2}}}dt
\end{equation}
and
\begin{equation}
V(\phi)=\sqrt{3B^2\beta^2 t^{2\beta-2}-\rho_{0}x_{8}}\times
\sqrt{3B^2\beta^2
t^{2\beta-2}+2B\beta(\beta-1)t^{\beta-2}+\rho_{0}\omega_{b}x_{8}}
\end{equation}

\begin{figure}
\includegraphics[height=3in]{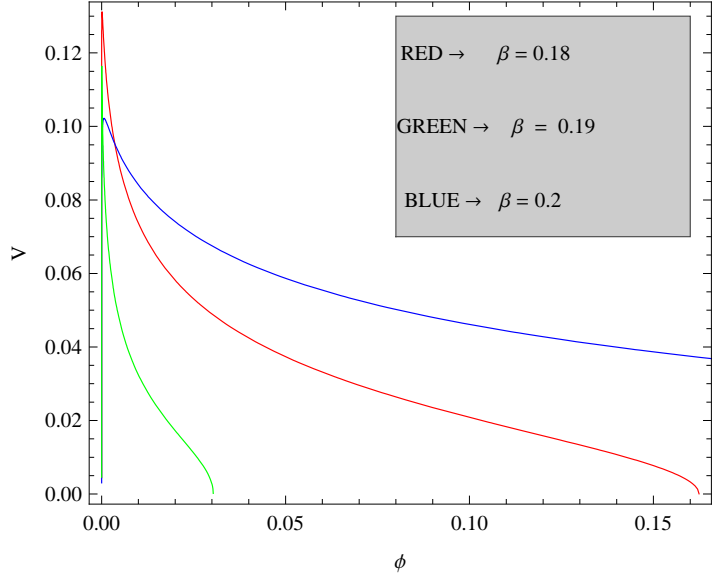}~~~~
\caption{The variation of $V$ against $\phi$ from (100) and (101)
for $B = 1,\omega_{b}=.3,\rho_{0}=5,\delta=.2$ and $\beta=0.18,
0.19, 0.2$} \vspace{7mm}
\end{figure}

The slow-roll parameters will be,
\begin{equation}
\epsilon=2\left(\frac{\beta-1}{t}\right)^2\times
\frac{\rho_{0}x_{8}-3B^2\beta^2
t^{2\beta-2}}{2B\beta(\beta-1)t^{\beta-2}+\rho_{0}\omega_{b}x_{8}}
\end{equation}
and
\begin{eqnarray*}
\eta=\frac{2(\beta-1)(\beta-2)}{t^2}\times\frac{\rho_{0}x_{8}-3B^2\beta^2
t^{2\beta-2}}{2B\beta(\beta-1)t^{\beta-2}+\rho_{0}\omega_{b}x_{8}}-\left(\frac{\beta-1}{t}\right)\times
\end{eqnarray*}
\begin{equation}
\left(\frac{\rho_{0}x_{8}-3B^2\beta^2
t^{2\beta-2}}{2B\beta(\beta-1)t^{\beta-2}+\rho_{0}\omega_{b}x_{8}}\right)^2
\times\frac{\partial}{\partial{t}}\left[\frac{2B\beta(\beta-1)t^{\beta-2}+\rho_{0}\omega_{b}x_{8}}
{\rho_{0}x_{8}-3B^2\beta^2 t^{2\beta-2}}\right]
\end{equation}

\begin{figure}
\includegraphics[height=3.5in]{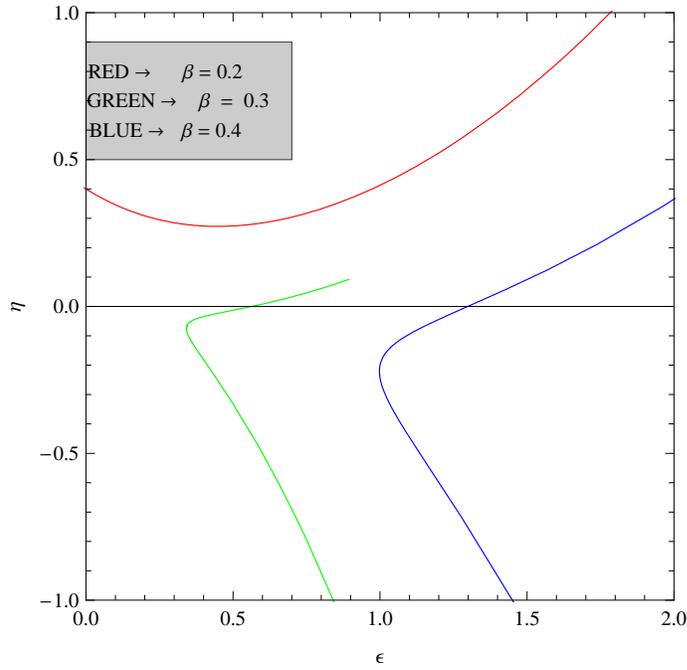}~~~~
\caption{The variation of $\eta$ against $\epsilon$ from (102) and
(103) for $B = 1,\omega_{b}=.3,\rho_{0}=5,\delta=.2$ and
$\beta=0.2, 0.3, 0.4$} \vspace{7mm}
\end{figure}

Fig. 19 shows the variation of $V$ against $\phi$. It has been
seen that $V$ decreases as $\phi$ increases. Also fig. 20
describes the variation of $\eta$ against $\epsilon$.

\section{\normalsize\bf{Mixture of Generalized Chaplygin
Gas and Barotropic Fluid with Tachyonic Field}}

Let us consider the universe is filled with the mixture of
generalized Chaplygin Gas, barotropic fluid and tachyonic field.
This generalized Chaplygin Gas is considered a perfect fluid is
given by,
\begin{equation}
p_{c}=-C/{\rho}_{c}^{\gamma}~~,~~~ 0\le \gamma \le 1, C>0.
\end{equation}

and the EOS of the barotropic fluid is given by,
\begin{equation}
p_{b}=\omega_{b} \rho_{b}
\end{equation}

If the energy density of the fluid is a function of volume only,
the temperature of the fluid remains zero at any pressure or
volume, violating the third law of thermodynamics.  The total
energy density and pressure are respectively given by,
\begin{equation}
\rho_{tot}=\rho_{c}+\rho_{b}+\rho_{t}
\end {equation}
and
\begin{equation}
p_{tot}=p_{c}+p_{b}+p_{t}
\end {equation}
where $p_{c}$ and $\rho_{c}$ are the pressure and density of the
generalized Chaplygin gas respectively and $p_{b}$ and $\rho_{b}$
are the pressure and density of the barotropic fluid respectively
and $p_{t}$ and $\rho_{t}$ are the pressure and density of the
tachyonic field respectively. Now we consider two possible states:
(i) without interaction and
(ii) with interaction.\\

\subsection{\normalsize\bf{Without Interaction}}
 The energy conservation equation is,
\begin{equation}
\dot{\rho}_{tot}+3\frac{\dot{a}}{a}(\rho_{tot}+p_{tot})=0
\end{equation}
Suppose the fluids do not interact with each other. Then the above
equation may be written as,
\begin{equation}
\dot{\rho}_{c}+3\frac{\dot{a}}{a}(\rho_{c}+p_{c})=0
\end{equation}
\begin{equation}
\dot{\rho}_{b}+3\frac{\dot{a}}{a}(\rho_{b}+p_{b})=0
\end{equation}
and
\begin{equation}
\dot{\rho}_{t}+3\frac{\dot{a}}{a}(\rho_{t}+p_{t})=0
\end{equation}

Now from equations (104) and (109), after eliminating $p_{c}$ we
get $\rho_{c}$ in terms of the scale factor,
\begin{equation}
\rho_{c}=\left[C+\rho_{c}'
a^{-3(1+\gamma)}\right]^{\frac{1}{1+\gamma}}
\end{equation}
and from (105) and (110) we get,
\begin{equation}
\rho_{b}=\rho_{b}'~a^{-3(1+\omega_{b})}
\end{equation}
where $\rho_{c}'$ and $\rho_{b}'$ are the integrating constants.\\

From (106), (112) and (113) we get,
\begin{equation}
\rho_{t}=3H^2-\left[C+\rho_{c}'
a^{-3(1+\gamma)}\right]^{\frac{1}{1+\gamma}}-\rho_{b}'~a^{-3(1+\omega_{b})}
\end{equation}
Thus from (107),(111) and (114) we get,
\begin{equation}
p_{t}=-3H^2-2\dot{H}+C\left[C+\rho_{c}'
a^{-3(1+\gamma)}\right]^{\frac{-\gamma}{1+\gamma}}-\omega_{b}\rho_{b}'~a^{-3(1+\omega_{b})}
\end{equation}
\\

{\normalsize{\bf{Case I:}}}\\

In the case of {\bf{Logamediate Inflation}} the energy density and
the pressure of the tachyonic fluid becomes,
\begin{equation}
\rho_{t}=\frac{3A^{2}\alpha^2 (\ln t)^{2\alpha-2}}{t^{2}}
-[C+\rho_{c}'x_{1}]^{\frac{1}{(1+\gamma)}}-\rho_{b}'x_{5}
\end{equation}
\begin{equation}
p_{t}=-\frac{3A^{2}\alpha^2 (\ln t)^{2\alpha-2}}{t^{2}}
+\frac{2A\alpha(\ln t -\alpha+1)(\ln
t)^{\alpha-2}}{t^2}+C[C+\rho_{c}'x_{1}]^{\frac{-\gamma}{(1+\gamma)}}-\rho_{b}'\omega_{b}x_{5}
\end{equation}
where, $x_{1}=\exp (-3A(1+\gamma)(\ln t)^{\alpha})$ and
$x_{5}=\exp (-3A(1+\omega_{b})(\ln t)^{\alpha})$.\\

From the equations (27), (28), (116) and (117), the tachyonic
field and the tachyonic potential are obtained as,
\begin{equation}
\phi=\int\sqrt{\frac{\frac{2A\alpha(-\ln t + \alpha-1)(\ln
t)^{\alpha-2}}{t^{2}}+\rho_{c}'x_{1}[C+\rho_{c}'x_{1}]^\frac{-\gamma}{\gamma+1}+\rho_{b}'(1+\omega_{b})x_{5}}
{{[C+\rho_{c}'x_{1}]^\frac{1}{\gamma+1}+\rho_{b}'x_{5}}-\frac{3A^2\alpha^2(\ln
t)^{2\alpha-2}}{t^2}}}~dt
\end{equation}
and
\begin{eqnarray*}
V(\phi)=\sqrt{\frac{3A^{2}\alpha^2 (\ln t)^{2\alpha-2}}{t^{2}}
-[C+\rho_{c}'x_{1}]^{\frac{1}{(1+\gamma)}}-\rho_{b}'x_{5}}~~\times~~~~~~~~~~~~~~~~~~~~~~~~~~
\end{eqnarray*}
\begin{equation}
\sqrt{\frac{3A^{2}\alpha^2 (\ln t)^{2\alpha-2}}{t^{2}}
-\frac{2A\alpha(\ln t -\alpha+1)(\ln
t)^{\alpha-2}}{t^2}-C[C+\rho_{c}'x_{1}]^{\frac{-\gamma}{(1+\gamma)}}+\rho_{b}'\omega_{b}x_{5}}
\end{equation}

From (7), (8), (29), (30) and (118) we get the slow-roll
parameters,
\begin{equation}
\epsilon=2\left(\frac{\alpha-1-\ln t}{t\ln t}\right)^2\times
\frac{{[C+\rho_{c}'x_{1}]^\frac{1}{\gamma+1}+\rho_{b}'x_{5}}-\frac{3A^2\alpha^2(\ln
t)^{2\alpha-2}}{t^2}}{\frac{2A\alpha(-\ln t + \alpha-1)(\ln
t)^{\alpha-2}}{t^{2}}+\rho_{c}'x_{1}[C+\rho_{c}'x_{1}]^\frac{-\gamma}{\gamma+1}+\rho_{b}'(1+\omega_{b})x_{5}}
\end{equation}
and
\begin{eqnarray*}
\eta=2\times\frac{{[C+\rho_{c}'x_{1}]^\frac{1}{\gamma+1}+\rho_{b}'x_{5}}-\frac{3A^2\alpha^2(\ln
t)^{2\alpha-2}}{t^2}}{\frac{2A\alpha(-\ln t + \alpha-1)(\ln
t)^{\alpha-2}}{t^{2}}+\rho_{c}'x_{1}[C+\rho_{c}'x_{1}]^\frac{-\gamma}{\gamma+1}+\rho_{b}'(1+\omega_{b})x_{5}}
 \times\left(\frac{2(\ln t)^2-3(\alpha-1)\ln
t+(\alpha-1)(\alpha-2)}{t^2 (\ln t)^2}\right)-
\end{eqnarray*}
\begin{eqnarray*}
\left(\frac{\alpha-1-\ln t}{t\ln t}\right)
\left(\frac{{[C+\rho_{c}'x_{1}]^\frac{1}{\gamma+1}+\rho_{b}'x_{5}}-\frac{3A^2\alpha^2(\ln
t)^{2\alpha-2}}{t^2}}{\frac{2A\alpha(-\ln t + \alpha-1)(\ln
t)^{\alpha-2}}{t^{2}}+\rho_{c}'x_{1}[C+\rho_{c}'x_{1}]^\frac{-\gamma}{\gamma+1}+\rho_{b}'(1+\omega_{b})x_{5}}
\right)^2
\end{eqnarray*}
\begin{equation}
\frac{\partial}{\partial{t}}\left[\frac{\frac{2A\alpha(-\ln t +
\alpha-1)(\ln
t)^{\alpha-2}}{t^{2}}+\rho_{c}'x_{1}[C+\rho_{c}'x_{1}]^\frac{-\gamma}{\gamma+1}+\rho_{b}'(1+\omega_{b})x_{5}}
{{[C+\rho_{c}'x_{1}]^\frac{1}{\gamma+1}+\rho_{b}'x_{5}}-\frac{3A^2\alpha^2(\ln
t)^{2\alpha-2}}{t^2}}\right]
\end{equation}

{\normalsize{\bf{Case II:}}}\\

In case of {\bf{Intermediate Scenario}}, the energy density and
the pressure of the tachyonic fluid is,
\begin{equation}
\rho_{t}=3B^2\beta^2
t^{2\beta-2}-[C+\rho_{c}'x_{2}]^{\frac{1}{(1+\gamma)}}-\rho_{b}'x_{6}
\end{equation}
and
\begin{equation}
p_{t}=-3B^2\beta^2 t^{2\beta-2}-2B\beta(\beta-1)t^{\beta-2}
+C[C+\rho_{0}x_{2}]^{\frac{-\gamma}{(1+\gamma)}}-\rho_{b}'\omega_{b}x_{6}
\end{equation}
where, $x_{2}=\exp (-3B(1+\gamma)t^\beta)$ and $x_{6}=\exp(-3B(1+\omega_{b})t^\beta)$. \\

Thus the tachyonic field and the tachyonic potential are obtained
as,
\begin{equation}
\phi=\int\sqrt{\frac{2B\beta
(\beta-1)t^{\beta-2}-C[C+\rho_{c}'x_{2}]^{\frac{-\gamma}{(1+\gamma)}}+\rho_{b}'(1+\omega_{b})x_{6}}{[C+\rho_{c}'x_{2}]
^{\frac{1}{(1+\gamma)}}+\rho_{b}'x_{6}-3B^2\beta^2
t^{2\beta-2}}}dt
\end{equation}
and
\begin{eqnarray*}
V(\phi)=\sqrt{3B^2\beta^2
t^{2\beta-2}-[C+\rho_{c}'x_{2}]^{\frac{1}{(1+\gamma)}}-\rho_{b}'x_{6}}~~~\times
\end{eqnarray*}
\begin{equation}
\sqrt{-3B^2\beta^2 t^{2\beta-2}-2B\beta(\beta-1)t^{\beta-2}
+C[C+\rho_{0}x_{2}]^{\frac{-\gamma}{(1+\gamma)}}-\rho_{b}'\omega_{b}x_{6}}
\end{equation}

From (16), (17), (29), (30) and (124) we get the slow-roll
parameters,
\begin{equation}
\epsilon=2\left(\frac{\beta-1}{t}\right)^2\times
\frac{[C+\rho_{c}'x_{2}]
^{\frac{1}{(1+\gamma)}}+\rho_{b}'x_{6}-3B^2\beta^2
t^{2\beta-2}}{2B\beta
(\beta-1)t^{\beta-2}-C[C+\rho_{c}'x_{2}]^{\frac{-\gamma}{(1+\gamma)}}+\rho_{b}'(1+\omega_{b})x_{6}}
\end{equation}
and
\begin{eqnarray*}
\eta=\frac{2(\beta-1)(\beta-2)}{t^2}\times\frac{[C+\rho_{c}'x_{2}]
^{\frac{1}{(1+\gamma)}}+\rho_{b}'x_{6}-3B^2\beta^2
t^{2\beta-2}}{2B\beta
(\beta-1)t^{\beta-2}-C[C+\rho_{c}'x_{2}]^{\frac{-\gamma}{(1+\gamma)}}+\rho_{b}'(1+\omega_{b})x_{6}}-
\end{eqnarray*}
\begin{eqnarray*}
\left(\frac{\beta-1}{t}\right) \left(\frac{[C+\rho_{c}'x_{2}]
^{\frac{1}{(1+\gamma)}}+\rho_{b}'x_{6}-3B^2\beta^2
t^{2\beta-2}}{2B\beta
(\beta-1)t^{\beta-2}-C[C+\rho_{c}'x_{2}]^{\frac{-\gamma}{(1+\gamma)}}+\rho_{b}'(1+\omega_{b})x_{6}}\right)^2
\end{eqnarray*}
\begin{equation}
\frac{\partial}{\partial{t}}\left[\frac{2B\beta
(\beta-1)t^{\beta-2}-C[C+\rho_{c}'x_{2}]^{\frac{-\gamma}{(1+\gamma)}}+\rho_{b}'(1+\omega_{b})x_{6}}{[C+\rho_{c}'x_{2}]
^{\frac{1}{(1+\gamma)}}+\rho_{b}'x_{6}-3B^2\beta^2
t^{2\beta-2}}\right]
\end{equation}

\subsection{\normalsize\bf{With Interaction}}

Now we consider an interaction between the tachyonic fluid, GCG
and barotropic fluid by introducing an interaction terms as a
product of the Hubble parameter and the energy densities of the
Chaplygin gas and barotropic fluid. Thus there
is an energy flow between the three fluids. \\

Now the equations of motion corresponding to the tachyonic field,
GCG and barotropic fluid are respectively,
\begin{equation}
\dot{\rho}_{t}+3\frac{\dot{a}}{a}(\rho_{t}+p_{t})=-3H\delta
\rho_{c}-3H\delta' \rho_{b}
\end{equation}
\begin{equation}
\dot{\rho}_{c}+3\frac{\dot{a}}{a}(\rho_{c}+p_{c})=3H\delta
\rho_{c}
\end{equation}
and
\begin{equation}
\dot{\rho}_{b}+3\frac{\dot{a}}{a}(\rho_{b}+p_{b})=3H\delta'
\rho_{b}
\end{equation}
where $\delta$ and $\delta'$ are the coupling constant.\\

From (104) and (129) we get,
\begin{equation}
\rho_{c}=\left[\frac{C}{1-\delta}+\rho_{c}''
a^{-3(1+\gamma)(1-\delta)}\right]^{\frac{1}{(1+\gamma)}}
\end{equation}
and from (105) and (130) we get,
\begin{equation}
\rho_{b}=\rho_{b}''~a^{-3(1+\omega_{b}-\delta')}
\end{equation}
where $\rho_{b}''$ and $\rho_{c}''$ are integrating constant.\\

{\normalsize{\bf{Case I:}}}\\

In case of {\bf{Logamediate Scenario}}, from
(21),(106),(128),(131),(132) we get the solutions:
\begin{equation}
\rho_{t}=\frac{3A^{2}\alpha^2 (\ln
t)^{2\alpha-2}}{t^{2}}-\left[\frac{C}{1-\delta}+\rho_{c}''
x_{3}\right]^{\frac{1}{(1+\gamma)}}-\rho_{b}''x_{7}
\end{equation}
where $x_{3}=\exp (-3A(1-\delta)(1+\gamma)(\ln t)^{\alpha})$ and
$x_{7}=\exp (-3A(1+\omega_{b}-\delta')(\ln t)^{\alpha})$. Hence
the pressure of the tachyonic field becomes,
\begin{equation}
p_{t}=-\frac{3A^{2}\alpha^2 (\ln t)^{2\alpha-2}}{t^{2}}
+\frac{2A\alpha(\ln t-\alpha+1)(\ln t)^{\alpha-2}}{t^2}
+C\left[\frac{C}{1-\delta}+\rho_{0}x_{3}\right]^{\frac{-\gamma}{(1+\gamma)}}-\rho_{b}''\omega_{b}x_{7}
\end{equation}

Solving the equations, the tachyonic field is obtained as,
\begin{equation}
\phi=\int\sqrt{\frac{\frac{2A\alpha(\ln t-\alpha+1)(\ln
t)^{\alpha-2}}{t^2}-\left(\frac{C\delta}{1-\delta}+\rho_{c}''x_{3}\right)\left(
\frac{C}{1-\delta}+\rho_{c}''x_{3}\right)^\frac{-\gamma}{1+\gamma}-\rho_{b}''(1+\omega_{b})x_{7}}{\frac{3A^{2}\alpha^2
(\ln t)^{2\alpha-2}}{t^{2}}-\left[\frac{C}{1-\delta}+\rho_{c}''
x_{3}\right]^{\frac{1}{(1+\gamma)}}-\rho_{b}''x_{7}}}~dt
\end{equation}
Also the potential will be of the form,\\
\begin{eqnarray*}
V(\phi)=\sqrt{\frac{3A^{2}\alpha^2 (\ln
t)^{2\alpha-2}}{t^{2}}-\left[\frac{C}{1-\delta}+\rho_{c}''
x_{3}\right]^{\frac{1}{(1+\gamma)}}-\rho_{b}''x_{7}}~~\times
\end{eqnarray*}
\begin{equation}
\sqrt{\frac{3A^{2}\alpha^2 (\ln t)^{2\alpha-2}}{t^{2}}
-\frac{2A\alpha(\ln t-\alpha+1)(\ln t)^{\alpha-2}}{t^2}
-C\left[\frac{C}{1-\delta}+\rho_{c}''x_{3}\right]^{\frac{-\gamma}{(1+\gamma)}}+\rho_{b}''\omega_{b}x_{7}}
\end{equation}

The slow-roll parameters are obtained as
\begin{equation}
\epsilon=2\left(\frac{\alpha-1-\ln t}{t\ln t}\right)^2\times
\frac{\frac{3A^{2}\alpha^2 (\ln
t)^{2\alpha-2}}{t^{2}}-\left[\frac{C}{1-\delta}+\rho_{c}''
x_{3}\right]^{\frac{1}{(1+\gamma)}}-\rho_{b}''x_{7}}{\frac{2A\alpha(\ln
t-\alpha+1)(\ln
t)^{\alpha-2}}{t^2}-\left(\frac{C\delta}{1-\delta}+\rho_{c}''x_{3}\right)\left(
\frac{C}{1-\delta}+\rho_{c}''x_{3}\right)^\frac{-\gamma}{1+\gamma}-\rho_{b}''(1+\omega_{b})x_{7}}
\end{equation}
and
\begin{eqnarray*}
\eta=\frac{\frac{3A^{2}\alpha^2 (\ln
t)^{2\alpha-2}}{t^{2}}-\left[\frac{C}{1-\delta}+\rho_{c}''
x_{3}\right]^{\frac{1}{(1+\gamma)}}-\rho_{b}''x_{7}}{\frac{2A\alpha(\ln
t-\alpha+1)(\ln
t)^{\alpha-2}}{t^2}-\left(\frac{C\delta}{1-\delta}+\rho_{c}''x_{3}\right)\left(
\frac{C}{1-\delta}+\rho_{c}''x_{3}\right)^\frac{-\gamma}{1+\gamma}-\rho_{b}''(1+\omega_{b})x_{7}}\times
\frac{4(\ln t)^2-6(\alpha-1)\ln t+2(\alpha-1)(\alpha-2)}{t^2 (\ln
t)^2}
\end{eqnarray*}
\begin{eqnarray*}
-~~\left(\frac{\alpha-1-\ln t}{t\ln t}\right)\times
\frac{\frac{3A^{2}\alpha^2 (\ln
t)^{2\alpha-2}}{t^{2}}-\left[\frac{C}{1-\delta}+\rho_{c}''
x_{3}\right]^{\frac{1}{(1+\gamma)}}-\rho_{b}''x_{7}}{\frac{2A\alpha(\ln
t-\alpha+1)(\ln
t)^{\alpha-2}}{t^2}-\left(\frac{C\delta}{1-\delta}+\rho_{c}''x_{3}\right)\left(
\frac{C}{1-\delta}+\rho_{c}''x_{3}\right)^\frac{-\gamma}{1+\gamma}-\rho_{b}''(1+\omega_{b})x_{7}}\times
\end{eqnarray*}
\begin{equation}
\frac{\partial}{\partial{t}}\left[\frac{\frac{2A\alpha(\ln
t-\alpha+1)(\ln
t)^{\alpha-2}}{t^2}-\left(\frac{C\delta}{1-\delta}+\rho_{c}''x_{3}\right)\left(
\frac{C}{1-\delta}+\rho_{c}''x_{3}\right)^\frac{-\gamma}{1+\gamma}-\rho_{b}''(1+\omega_{b})x_{7}}{\frac{3A^{2}\alpha^2
(\ln t)^{2\alpha-2}}{t^{2}}-\left[\frac{C}{1-\delta}+\rho_{c}''
x_{3}\right]^{\frac{1}{(1+\gamma)}}-\rho_{b}''x_{7}}\right]
\end{equation}

{\normalsize{\bf{Case II:}}}\\

In case of Intermediate Scenario, the energy density and the
pressure of the tachyonic fluid is,

\begin{equation}
\rho_{t}=3B^2\beta^2
t^{2\beta-2}-\left[\frac{C}{1-\delta}+\rho_{c}''x_{4}\right]^{\frac{1}{(1+\gamma)}}-\rho_{b}''x_{8}
\end{equation}

Hence
\begin{equation}
p_{t}=-3B^2\beta^2
t^{2\beta-2}-2B\beta(\beta-1)t^{\beta-2}-\rho_{c}''(1-\delta)(1+\gamma)
x_{4}\left[\frac{C}{1-\delta}+\rho_{c}''x_{4}\right]^{\frac{-\gamma}{(1+\gamma)}}-\rho_{b}''\omega_{b}x_{8}
\end{equation}
where $x_{4}=\exp (-3B(1-\delta)(1+\gamma)t^\beta)$ and
$x_{8}=\exp(-3B(1+\omega_{b}-\delta')t^\beta)$.

Thus the tachyonic field and the tachyonic potential are obtained
as,
\begin{equation}
\phi=\int\sqrt{\frac{2B\beta(\beta-1)t^{\beta-2}-\left[\frac{C}{1-\delta}+\rho_{c}''x_{4}\right]^{\frac{1}{(1+\gamma)}}
+\rho_{c}''(1-\delta)(1+\gamma)x_{4}\left[\frac{C}{1-\delta}+\rho_{c}''x_{4}\right]^{\frac{-\gamma}{(1+\gamma)}}+\rho_{b}''\omega_{b}x_{8}}
{\left[\frac{C}{1-\delta}+\rho_{c}''x_{4}\right]
^{\frac{1}{(1+\gamma)}}+\rho_{b}''x_{8}-3B^2\beta^2
t^{2\beta-2}}}~dt
\end{equation}
and
\begin{eqnarray*}
V(\phi)=\sqrt{3B^2\beta^2
t^{2\beta-2}-\left[\frac{C}{1-\delta}+\rho_{c}''x_{4}\right]^{\frac{1}{(1+\gamma)}}-\rho_{b}''x_{8}}~~\times
\end{eqnarray*}
\begin{equation}
\sqrt{3B^2\beta^2
t^{2\beta-2}+2B\beta(\beta-1)t^{\beta-2}+\rho_{c}''(1-\delta)(1+\gamma)
x_{4}\left[\frac{C}{1-\delta}+\rho_{c}''x_{4}\right]^{\frac{-\gamma}{(1+\gamma)}}+\rho_{b}''\omega_{b}x_{8}}
\end{equation}

The slow-roll parameters will be,
\begin{equation}
\epsilon=2\left(\frac{\beta-1}{t}\right)^2\times
\frac{\left[\frac{C}{1-\delta}+\rho_{c}''x_{4}\right]
^{\frac{1}{(1+\gamma)}}+\rho_{b}''x_{8}-3B^2\beta^2
t^{2\beta-2}}{2B\beta(\beta-1)t^{\beta-2}-\left[\frac{C}{1-\delta}+\rho_{c}''x_{4}\right]^{\frac{1}{(1+\gamma)}}
+\rho_{c}''(1-\delta)(1+\gamma)x_{4}\left[\frac{C}{1-\delta}+\rho_{c}''x_{4}\right]^{\frac{-\gamma}{(1+\gamma)}}+\rho_{b}''\omega_{b}x_{8}}
\end{equation}
and
\begin{eqnarray*}
\eta=\frac{2(\beta-1)(\beta-2)}{t^2}\times\frac{\left[\frac{C}{1-\delta}+\rho_{c}''x_{4}\right]
^{\frac{1}{(1+\gamma)}}+\rho_{b}''x_{8}-3B^2\beta^2
t^{2\beta-2}}{2B\beta(\beta-1)t^{\beta-2}-\left[\frac{C}{1-\delta}+\rho_{c}''x_{4}\right]^{\frac{1}{(1+\gamma)}}
+\rho_{c}''(1-\delta)(1+\gamma)x_{4}\left[\frac{C}{1-\delta}+\rho_{c}''x_{4}\right]^{\frac{-\gamma}{(1+\gamma)}}+\rho_{b}''\omega_{b}x_{8}}
\end{eqnarray*}
\begin{eqnarray*}
-\left(\frac{\beta-1}{t}\right)\times\left(\frac{\left[\frac{C}{1-\delta}+\rho_{c}''x_{4}\right]
^{\frac{1}{(1+\gamma)}}+\rho_{b}''x_{8}-3B^2\beta^2
t^{2\beta-2}}{2B\beta(\beta-1)t^{\beta-2}-\left[\frac{C}{1-\delta}+\rho_{c}''x_{4}\right]^{\frac{1}{(1+\gamma)}}
+\rho_{c}''(1-\delta)(1+\gamma)x_{4}\left[\frac{C}{1-\delta}+\rho_{c}''x_{4}\right]^{\frac{-\gamma}{(1+\gamma)}}+\rho_{b}''\omega_{b}x_{8}}
\right)^2\times
\end{eqnarray*}
\begin{equation}
\frac{\partial}{\partial{t}}\left[\frac{2B\beta(\beta-1)t^{\beta-2}-\left[\frac{C}{1-\delta}+\rho_{c}''x_{4}\right]^{\frac{1}{(1+\gamma)}}
+\rho_{c}''(1-\delta)(1+\gamma)x_{4}\left[\frac{C}{1-\delta}+\rho_{c}''x_{4}\right]^{\frac{-\gamma}{(1+\gamma)}}+\rho_{b}''\omega_{b}x_{8}}
{\left[\frac{C}{1-\delta}+\rho_{c}''x_{4}\right]
^{\frac{1}{(1+\gamma)}}+\rho_{b}''x_{8}-3B^2\beta^2
t^{2\beta-2}}\right]
\end{equation}

\section{\normalsize\bf{Discussions}}

In this work, we have considered a model of two and three
component mixture i.e., mixture of Chaplygin gas and barotropic
fluid with tachyonic field. In the case, when they have no
interaction then both of them retain their own properties. Let us
consider an energy flow between barotropic and tachyonic fluids.
In both the cases we find the exact solutions for the tachyonic
field and the tachyonic potential and show that the tachyonic
potential follows the asymptotic behavior. Here the tachyonic
field behaves as the dark energy component. For the tachyonic dark
matter, GCG is considered as a suitable dark energy model. Later
we have also considered an interaction between these two fluids by
introducing a coupling term. The coupling function decays with
time indicating a strong energy flow at the initial period and
weak stable interaction at later stage. To keep the observational
support of recent acceleration we have considered two particular
forms: (i) Logamediate Scenario (ii) Intermediate Scenario, of
evolution of the Universe. In both the scenarios, we have obtained
the expressions of statefinder parameters. We graphically show the
natures of statefinder parameters for evolution of the universe in
both the cases. We have considered the mixture of Chaplygin gas
and tachyonic field with and without interactions. Logamediate and
intermediate expansions have been considered with and without
interaction cases. For all possible cases we have obtained the
natures of potentials and slow-roll parameters graphically. Next
we have also considered the mixture of barotropic fluid and
tachyonic field with and without interactions. Logamediate and
intermediate expansions have been considered with and without
interaction cases also. For all possible cases we have also
obtained the natures of potentials and slow-roll parameters
graphically. Finally, we have considered the mixture of tachyonic
field, Chaplygin gas and barotropic fluid with and without
interactions. Logamediate and intermediate expansions have been
considered with and without interaction cases. For all possible
cases we have obtained the natures of potentials and slow-roll
parameters graphically. Thus the present work shows the natures of
statefinder and slow-roll parameters in both logamediate and
intermediate scenarios for the evolution of the universe.\\\\

{\bf References:}\\
\\
$[1]$  N. A. Bachall, J. P. Ostriker, S. Perlmutter and P. J.
Steinhardt, {\it Science} {\bf 284} 1481 (1999).\\
$[2]$ S. J. Perlmutter et al, {\it Astrophys. J.} {\bf 517} 565
(1999).\\
$[3]$ A. G. Riess et al, {\it Astron. J.} {\bf 116} 1009
(1998).\\
$[4]$ P. M. Garnavich et al, {\it Astrophys. J.} {\bf 509} 74
(1998).\\
$[5]$ G. Efstathiou et al, {\it astro-ph}/9812226.\\
$[6]$ B. Ratra and P. J. E. Peebles, {\it Phys. Rev. D} {\bf 37}
3406 (1988).\\
$[7]$ R. R. Caldwell, R. Dave and P. J. Steinhardt, {\it
Phys. Rev. Lett.} {\bf 80} 1582 (1998).\\
$[8]$ A. Kamenshchik, U. Moschella and V. Pasquier, {\it Phys.
Lett. B} {\bf 511} 265 (2001).\\
$[9]$ V. Gorini, A. Kamenshchik, U. Moschella and V. Pasquier,
{\it gr-qc}/0403062.\\
$[10]$ V. Gorini, A. Kamenshchik and U. Moschella, {\it Phys. Rev.
D} {\bf 67} 063509 (2003).\\
$[11]$ U. Alam, V. Sahni , T. D. Saini and A.A. Starobinsky, {\it
Mon. Not. Roy. Astron. Soc.} {\bf 344}, 1057 (2003).\\
$[12]$ M. C. Bento, O. Bertolami and A. A. Sen, {\it Phys. Rev. D}
{66} 043507 (2002).\\
$[13]$ H. B. Benaoum, {\it hep-th}/0205140.\\
$[14]$ U. Debnath, A. Banerjee and S. Chakraborty, {\it Class.
Quantum Grav.} {\bf 21} 5609 (2004).\\
$[15]$ A. Sen, {\it JHEP} {\bf04} 048 (2002); {\it JHEP} {\bf 07}
065 (2002); {\it Mod. Phys. Lett. A} {\bf 17}
1797(2002).\\
$[16]$ J. S. Bagla, H. K. Jassal and T. Padmanabhan, {\it Phys.
Rev. D} {\bf 67} 063504 (2003).\\
$[17]$ E. J. Copeland, M. R. Garousi, M. Sami and S. Tsujikawa,
{\it Phys. Rev D} {\bf 71} 043003 (2005).\\
$[18]$ G. Calcagni and A. R. Liddle, {\it astro-ph}/0606003.\\
$[19]$ E. J. Copeland, M. Sami and S. Tsujikawa, {\it Int. J.
Mod. Phys. D} {\bf 15} 1753 (2006).\\
$[20]$ A. DeBenedictis, A. Das and S. Kloster, {\it Gen. Rel.
Grav.} {\bf 36} 2481 (2004).\\
$[21]$ A. Das, S. Gupta, T. D. Saini and S. Kar, {\it Phys. Rev D}
{\bf 72} 043528 (2005).\\
$[22]$ T. Padmanabhan, {\it Phys. Rev. D} {\bf 66} 021301 {\bf R}
(2002).\\
$[23]$ M. Sami, {\it Mod. Phys. Lett. A} {\bf 18} 691 (2003).\\
$[24]$ B. C. Paul and D. Paul, {\it Int. J. Mod. Phys. D} {\bf 14}
1831 (2005).\\
$[25]$ J. D. Barrow and N. J. Nunes, {\it Phys. Rev. D} {\bf 76}
043501 (2007).\\
$[26]$ J. D. Barrow, {\it Phys. Lett. B} {\bf 235} 40 (1990); J.
D. Barrow and P. Saich, {\it Phys. Lett. B} {\bf 249} 406 (1990);
J. D. Barrow, A. R. Liddle, and C. Pahud, {\it Phys. Rev. D} {\bf
74} 127305 (2006).\\
$[27]$ V. Sahni, T. D. Saini, A. A. Starobinsky and U. Alam, {\it
JETP Lett.} {\bf 77} 201 (2003).\\

\end{document}